\documentclass[]{aastex631}

\usepackage{amsmath,amstext}
\usepackage[T1]{fontenc}
\usepackage[figure,figure*]{hypcap}
\usepackage{yfonts}
\usepackage{bbold}

\usepackage{algorithm}
\usepackage[noend]{algpseudocode}

\newcommand{\gsim}{\lower.7ex\hbox{$\;\stackrel{\textstyle>}{\sim}\;$}}
\newcommand{\lsim}{\lower.7ex\hbox{$\;\stackrel{\textstyle<}{\sim}\;$}}

\newcommand{\PS}{PS1}

\newcommand{\au}{au}

\def \MIDAS {Michigan Institute for Data and AI in Society, University of Michigan, Ann Arbor, MI 48109, USA}
\def \Physics {Department of Physics, University of Michigan, Ann Arbor, MI 48109, USA}

\shorttitle{Pan-STARRS1 Planet Search}

\begin{document}

\title{A Pan-STARRS Search for Distant Planets: Part 1}

\author[0000-0002-1139-4880]{Matthew~J.~Holman}
\affil{Harvard-Smithsonian Center for Astrophysics, 60 Garden St., MS 51, Cambridge, MA 02138, USA}
\correspondingauthor{Matthew~J.~Holman}
\email{mholman@cfa.harvard.edu}

\author[0000-0003-4827-5049]{Kevin~J.~Napier}
\affil{Harvard-Smithsonian Center for Astrophysics, 60 Garden St., MS 51, Cambridge, MA 02138, USA}
\affiliation{\MIDAS}
\affiliation{\Physics}

\author[0000-0001-5133-6303]{Matthew~J.~Payne}
\affil{Harvard-Smithsonian Center for Astrophysics, 60 Garden St., MS 51, Cambridge, MA 02138, USA}

\author[0009-0005-5452-0671]{Jacob A. Kurlander} 
\affiliation{DiRAC Institute and the Department of Astronomy, University of Washington, 3910 15th Ave NE, Seattle, WA 98195, USA} 

\accepted{23 May 2025}

\begin{abstract}
We present a search for distant planets in Pan-STARRS1. We calibrated our search by injecting an isotropic control population of synthetic detections into Pan-STARRS1 source catalogs, providing a high-fidelity alternative to injecting synthetic sources at the image level. We found that our method is sensitive to a wide range of distances, as well as all rates and directions of motion. We identified 692 solar system objects (109 of which are not yet listed in the Minor Planet Center's database), including 642 TNOs, 23 of which are dwarf planets.  By raw number of detections, this makes our search the third most productive Kuiper Belt survey to date, in spite of the fact that we did not explicitly search for objects closer than 80 au. Although we did not find Planet Nine or any other planetary objects, we were able to show that the remaining parameter space for Planet Nine is highly concentrated in the galactic plane.  

\end{abstract}


\section{Introduction}

A number of lines of evidence suggest that additional planets may remain to be discovered in our solar system. The process of planet formation is expected to have resulted in more planets than we observe today, and although most such objects would have been ejected, others may remain~\citep{Stern.1991,Chiang.2007,Silsbee.2018,Gladman.2021}. The orbits of some ``detached'' or ``extreme'' trans-Neptunian objects (TNOs), such as Sedna~\citep{Brown.2004} and 2012~VP113~\citep{Trujillo.2014}, with large perihelia ($q>65$) and semi-major axes that are interior to the inner Oort Cloud ($a<600$), cannot be readily explained by a combination of gravitational perturbations of the known planets and the galaxy, but they can be generated in simulations that include additional planets~\citep{Gladman.2002, Brown.2004, Trujillo.2014,Kavelaars.2020}.  An additional planet could account for the abundance of TNOs in distant mean motion resonances with Neptune~\citep{Huang.2022}, as well as for the warped plane of the Kuiper belt~\citep{Volk.2017} and other dynamical features of the trans-Neptunian region~\citep{Lykawka.2007,Lykawka.2008,Lykawka.2023}. A distant gas giant could generate the apparent apsidal alignment of extreme TNOs, those with large semi-major axes and perihelion distances~\citep{Trujillo.2014,Brown.2016,Batygin.2016,Sheppard.2016}.  

Searches for additional planets have been conducted, leading to the discovery of dwarf planets but no  larger bodies to date.  However, these surveys have not ruled out the existence of additional planets.  The existing surveys have  been sensitive to only the brightest objects~\citep{Brown.2005,Schwamb.2010,Rabinowitz.2012}, have been confined to specific regions of the sky where theoretical models suggest an additional planet is likely to be found~\citep{Brown.2024}, or have been restricted to particular regions of the sky~\citep{Gerdes.2016,Bernardinelli.2020,Bernardinelli.2022}.  (See \citet{Bannister.2020} and \citet{Gladman.2021} for thorough reviews of TNO surveys and results.)

The Vera Rubin Observatory's Legacy Survey of Space and Time (LSST) will be sensitive to moving objects brighter than magnitude $r\sim24.5$, even in distant orbits~\citep{Jones.2016,Trilling.2018}.  LSST will cover all of the southern hemisphere and most, if not all, of the ecliptic.  
With commissioning underway, it is tempting to simply wait for LSST to begin.  However, there are good reasons to exhaust other data sets.  A number of the surveys probe to fainter limiting magnitudes than LSST's wide field survey will.  Most of the northern hemisphere will not be searched by LSST, leaving a large volume of discovery space unexplored.

The Panoramic Survey Telescope and Rapid Response System 1 (Pan-STARRS1) has covered more than three quarters of the sky, including all of the northern sky and the entire ecliptic~\citep{Kaiser.2010, Chambers.2016}, which will complement LSST's expected coverage of the southern sky.
Pan-STARRS1 supported a wide variety of investigations, including searches for solar system objects.
Although the survey pattern was, and still is, optimized for detecting near-Earth objects (NEOs)~\citep{Denneau.2013}, the survey's moving object pipeline is sensitive to objects moving as slowly as $\sim0.05\deg/\mathrm{day}$, making Pan-STARRS1 one of the most prolific systems for the discovery of Centaurs. 
However, a number of challenges must be overcome to identify more distant bodies in Pan-STARRS1 data.  

The motion of distant objects can be too slow to confidently discern with a single night of Pan-STARRS1 data.
The Pan-STARRS1 observations within a night of a given region of sky typically span only 20-60~minutes.  Particularly distant TNOs, such as Sedna at ${\sim}87$~au during the survey, might have an apparent motion of ${\sim}1.5\arcsec/\mathrm{hr}$ or less at opposition, yielding a total motion of ${\sim}0.5-1.5\arcsec$.  With an astrometric uncertainty as large as ${\sim}0.2-0.3\arcsec$, that amount of motion might be only marginally detected.  
Thus,  data from multiple nights must be combined to make unambiguous discoveries of distant outer solar system objects with Pan-STARRS1.

The Pan-STARRS1 camera's fill factor is comparatively low.  There are both inter-chip and intra-chip gaps.  The dead space in the focal plane is distributed on a range of scales, neither uniformly distributed across the field of view nor concentrated in large contiguous areas.  Thus, it is difficult to estimate {\it a priori} the likelihood that a sequence of exposures will result in the detection of a moving object~\citep{Denneau.2013}.

The Pan-STARRS1 survey pattern
was not designed for the characterization of the survey's observational biases and efficiency for discovering TNOs or more distant objects.
Most well-characterized TNO surveys define sets of  ``discovery fields''~\citep{Petit.2011,Bannister.2016, Trilling.2024}. These are typically large, contiguous patches of sky that are repeatedly observed on a carefully-chosen sequence of nights.  The arrangement of the fields and the timing of the observations are selected such that most of the objects discovered in those fields will automatically receive sufficient observations to allow the objects' future sky-plane locations to be predicted accurately enough to be observed with telescopes and cameras with smaller fields of view.  If all of the moving objects identified in the discovery fields are followed up, one can isolate the detection efficiency and observational biases to just those fields. Such well-designed surveys allow for a clean separation between the discovery phase and the follow-up and characterization phases.
However, the Pan-STARRS1 survey cannot be easily divided into discovery and follow-up/characterization observations.  Instead, nearly all of the observations must be used to support a combination of discovery and characterization, similar to the recent TNO searches with data from the {\it Dark Energy Survey}~\citep{Gerdes.2016,Khain.2020,Bernardinelli.2020,Bernardinelli.2022}.

Finally, the rate of false detections in Pan-STARRS1 is enormously high~\citep{Denneau.2013}, which means that more data points are needed for a confident discovery.  However, for much of the survey, the observations of a particular region of the sky were concentrated within a single lunation each year, with no year receiving enough observations to make a definitive discovery.  In those cases, observations from multiple years must be combined to obtain enough data to discover distant solar system objects.
Linking sparsely sampled observations across large time spans, in the presence of an overwhelming background of spurious detections presents a huge computational problem that has demanded the development of novel algorithms to efficiently address this problem~\citep{Holman.2018a, Holman.2018b}.

Despite these challenges, this rich data set has enabled the discovery of hundreds of TNOs~\citep{Weryk.2016,Holman.2018b}, including Neptune Trojans~\citep{Lin.2016}, a TNO in a retrograde orbit~\citep{Chen.2016}, and a dwarf planet~\citep{Holman.2018b}.

What has been missing from these previous investigations is a systematic means of determining the observational biases that result from the Pan-STARRS1 survey pattern and the characteristics of its camera.  Without that, we cannot, for example, establish if we have discovered most members of a population or a small fraction of a much larger population.

The gold standard for characterizing solar system searches is to inject PSF-matched sources, with appropriate noise, directly into the images, at the locations predicted by models of the sky-plane motion of a sample of objects.  The images are then processed and searched, with the identity of the synthetic sources   revealed only at the end.  This is often repeated in order to boost the statistical significance of the results.  Injecting synthetic sources directly into the images naturally accounts for the effects of stellar crowding, defects in the images, changes in the image quality and transparency, and a host of other effects that might not be easily identified in advance.   Injection-recovery experiments have been used to not only determine the detection efficiency as a function of source magnitude and rate of motion but to statistically test models for the size and orbital distributions of small body populations~\citep{Gladman.2001,Bernstein.2004,Lawler.2018,Napier.2024a}.  

Unfortunately, injecting synthetic sources directly into Pan-STARRS1 images and then processing and searching the implanted data is not feasible. The volume of images is so large that processing them involves large computing clusters, vast amounts of storage, and significant IT support.  This is done on a schedule, with planned data releases.  Repeating this process to build up statistics would be even more intractable.

If we had a means of determining the likelihood that a candidate synthetic source, of a given brightness, sky-plane location, and focal plane position, would be detected and recorded in the source catalogs, this would be essentially equivalent to injecting synthetic sources into the images.  And it would be far more manageable computationally.  But doing this properly would require careful attention to the focal plane configuration, the properties of the individual detectors, and the effects of bright stars.  Reliable models for the detection efficiency, as well as the astrometric and photometric uncertainties, would also be needed. 
Fortunately, all of these ingredients are provided, on a per-exposure basis, by the Pan-STARRS Image Processing Pipeline~\citep{Magnier.2013,Magnier.2020a,Magnier.2020b,Magnier.2020c}.

The survey we present here has much in common with the recent Centaur search of \citet{Kurlander.2025}.  The same data were used, as were a number of the processing routines, and the overall approach of injecting sources into the catalogs.  Aside from the search targets (Centaurs versus distant planets), the primary differences were in the linking algorithm and in the resulting survey simulators.  As it was focused on faster moving objects, for which the sky-plane angular rates of motion are measured with better relative precision, the \citet{Kurlander.2025} used the HelioLinC3D~\citep{Heinze.2022, Heinze.2023} version of the HelioLinC algorithm~\citep{Holman.2018a}.  For slower-moving targets, objects for which very little motion is apparent within a night, we needed to develop and use a different but related technique (section~\ref{SECN:PIPELINE}).  For their survey simulator, \citet{Kurlander.2025} fit a selection function to their synthetic objects as a function of orbits and magnitude.  As an alternative approach to survey simulation, we use a machine learning model to estimate the probability that we would have recovered an object in our data, given a vector of the epochs at which it would have appeared in our detection catalogs (section~\ref{sec:simulator}).

An alternative to injection-recovery is to use the known asteroids to characterize the search.  The vast majority of known asteroids have well determined orbits and reasonably accurate absolute magnitudes.  It is therefore possible to reliably predict the sky-plane location and expected magnitude of these objects in any given Pan-STARRS1 exposure.  Because there can be hundreds of asteroids in an individual Pan-STARRS1 exposure, the fraction of objects detected as a function of magnitude can be accurately estimated.  \citet{Brown.2024} used this approach in a calibrated search of publicly available Pan-STARRS1 data for ``Planet Nine''.  Although this method is effective, it does have limitations.  One can only calibrate the search where the sky-plane density of asteroid is high enough to permit a meaningful estimate of the detection efficiency.  Because there are relatively few asteroids with very high inclinations, a calibrated search of regions at high ecliptic latitudes requires a different method.  Similarly, the apparent rates of motion of main belt asteroids is much higher than that of TNOs or more distant objects, so care must be taken to ensure that the search does not depend upon the sky-plane rate of motion, having multiple detections in a night, or any other characteristic that might differ between the dynamical classes of objects.  Although there is a significant population of TNOs that straddle the Pan-STARRS1 detection limits, it is still small compared to the number of synthetic sources that is typically used to characterize a survey.

In the remainder of this paper we develop and test this approach for Pan-STARRS1 data and apply it to a search for distant planets.  In section~\ref{SECN:PS1}, we describe the Pan-STARRS1 system
and its Image Processing Pipeline (IPP) 
in more detail.
The generation of the orbits of our control population is described in section~\ref{SECN:ORBITS}.
In section~\ref{SECN:Filtering}, we detail the steps to identify and eliminate detections that are unlikely to correspond to real or synthetic moving objects, thereby speeding up our search.   In section~\ref{SECN:PIPELINE}, we describe our search algorithm.   We provide a qualitative analysis of our results in section~\ref{SECN:QA}.  We describe a novel approach to developing a survey simulator, based on our search results, in section~\ref{sec:simulator}.  In section~\ref{sec:p9}, we apply our survey simulator to the \citep{Brown.2021} sample of Planet Nine orbits.  We summarize our work and outline areas of future work in section~\ref{SECN:Conclusions}.

\section{Pan-STARRS1}
\label{SECN:PS1}

Pan-STARRS1, located on Hale\'akala, Hawaii, is an innovative wide-field optical system designed for multi-epoch, multi-color surveys~\citep{Hodapp.2004, Kaiser.2010,Tonry.2012,Schlafly.2012,Magnier.2013,Magnier.2020a,Magnier.2020b,Magnier.2020c}. 
 The 1.8~m aperture telescope is equipped with the GigaPixel Camera 1 (GPC1), consisting of an array of 60 CCD detectors, each 4800 pixels on a side.  These CCDs are prototype orthogonal transfer arrays (OTAs), with an $8\times8$ grid of subregions that can be read out independently.  GPC1 images a 7 sq. degree field of view~\citep{Onaka.2008, PS1_GPCa}.   

 Much of the data are taken in the PS1 g, r, i, z, and y filters~\citep{Tonry.2012} and are photometrically calibrated to better than 1\% accuracy \citep{Schlafly.2012, Tonry.2012b, Finkbeiner.2016}.  In addition, a large fraction of the data are taken in the w filter (400-820~nm).  The w-band exposures are less well calibrated, but are reliable at the 5-10\% level~\citep{Denneau.2013}.
 The exposure times are $\sim30-60$~sec.  Fields are typically imaged two to four times in a night, with successive exposures of the same field separated by $\sim 15-20$~minutes.  

Our search includes data from 2009 April 12 through 2017 Nov 1.  This spans the commissioning phase,  Pan-STARRS1 Science Consortium (PS1SC) surveys, and 3.5 subsequent years during which  $\sim90\%$ of the Pan-STARRS1 telescope time was dedicated to the search for NEOs.  
We exclude the y band data, which is particularly shallow.  Also, we exclude the data from the Pan-STARRS1 Medium Deep Survey, which covered ten selected fields (see Figure~1) with nightly observations, with longer exposures in the g, r, i, z, and y filters.  The cadence and depth of these exposures are distinct enough from the rest of the Pan-STARRS1 data that they complicate our analysis while adding little sky coverage. 

\begin{figure}[htp]
    \centering
    \includegraphics[trim = 0mm 0mm 0mm 0mm, clip, angle=0, width=0.5\columnwidth]{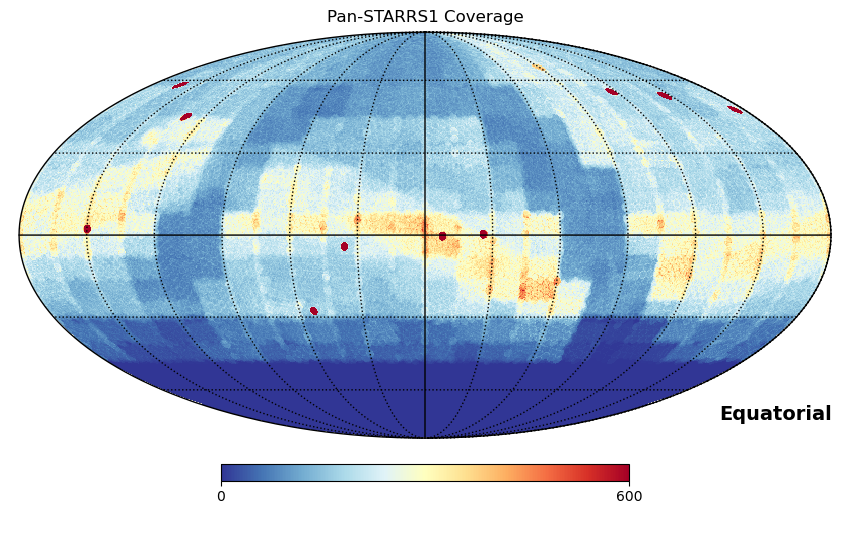}

    \caption{The coverage of the Pan-STARRS1 data used in this search, from 2009 April 2 until 2017 November, shown in a Mollweide projection in equatorial coordinates (RA 0 deg is at the center of the plot, with RA increasing to the left). The colors encode, with a linear scale, the number of exposures that cover the region, assuming a circular field of view with a $1.6\deg$ angular radius, ignoring the $\sim68\%$ fill factor. The ten Medium Deep fields are shown for completeness. 
    (Note that we skip any data from the Medium Deep fields, as well as from a few exposures for which the processing seems to have failed.) Each has thousands of exposures; we cap the maximum at 600 to keep those fields from dominating the color scale.    The median number of exposures is ${\sim}150$.  A fraction of the sky was not observed at all, and a few regions were observed more than ${\sim400}$ times.  The center of the plot is 0~hr and 0 
    }
    \label{FIG:COVERAGE}
    \end{figure}

We assembled the data from $708,554$ unique Pan-STARRS1 exposures for this search.
Figure~\ref{FIG:COVERAGE} shows the distribution of these exposures on the sky. 
These data covered $\sim80\%$ of the sky
over the course of 8 years.  Aside from the galactic plane and the regions too far south to be observed with Pan-STARRS1, most regions have 100-200 exposures; some regions of ecliptic have more than 400 exposures.  

    %

As noted earlier, the images collected by Pan-STARRS1 are processed by the IPP~\citep{Magnier.2013,Magnier.2020a,Magnier.2020b,Magnier.2020c}.   Although the IPP supports image registration, stacking, and subtraction, we adopted the catalogs of sources detected 
 in the direct, unsubtracted exposures as our primary data.  Those data are typically deeper than those from difference imaging.  The Pan-STARRS1 source detection data from each exposure are stored in a multi-extension FITS format.  In addition to detection catalogs, the files include detection efficiency data and other content that enables us to develop a high fidelity process for injecting sources into the detection catalogs.

\section{Control Population}
\label{SECN:ORBITS}

We determine the magnitude limit and completeness of our search by using a control population whose orbits and physical characteristics are known.  We generate synthetic detections for each of the Pan-STARRS1 exposures in our survey in a way that mimics the real source detection process.  Then we process and search the combined sample of real and synthetic detections.

Our control population needs to span the likely boundaries of detectability to allow us to identify and quantify those limits.  With the following procedure for generating trial orbits and absolute magnitudes we can easily control and sample the properties that most affect detectability.  First, we generate a random vector on the unit sphere to specify the barycentric direction to the object.  Next, we generate another random unit vector that is perpendicular to the first.  This specifies the orbit normal.  
We then select a barycentric distance $r$ uniformly between $25~\au$ and $1500~\au$ and scale the first unit vector by that value. Finally, we select an orbital eccentricity $e$ uniformly between 0 and $e_{max}=0.99$ and an orbital mean anomaly $\mathrm{M}$ uniformly between 0 and $2\pi$ radians.  This process uniquely specifies the six orbital elements.  We verify that the barycentric position vectors and orbit normals are uniformly distributed on the sky, the resulting longitudes of ascending node $\Omega$ and arguments of pericenter $\omega$ are uniformly distributed between 0 and $2\pi$ radians, and the inclinations $i$ are isotropic.  We adopt JD~2454466.5 TDB, 2008 Jan 1, as the epoch of the orbits.

Along with the orbital elements, we assign an absolute magnitude and color to each object. Our objective is to generate a distribution of absolute magnitudes that results in apparent magnitudes that span the detection limits for a wide range of distances.
We choose an apparent w-band magnitude uniformly between $w=19.5$ and $w=23.5$.  This spans the expected limiting magnitude of $w\sim22.5$ while allowing a large enough sample of fainter sources to measure the limiting magnitude and a large enough sample of brighter sources to measure the dependence on sky plane and focal plane location.  To model the change in apparent brightness with time, we convert the reference magnitude to an absolute magnitude with the approximation $H_w = w - 10\log_{10}(d)$, where $d$ is the barycentric distance.   To convert absolute magnitudes to apparent magnitudes we use the two parameter H-G system~\citep{Bowell.1989}, assuming $G=0.15$.
Following the approach of \citet{Bernardinelli.2022}, we select a $g-r$ color uniformly between 0 and 1.5. 
 Then we obtain the other colors with $g-i=\alpha_i(g-r) + \beta_i$, and $g-z=\alpha_z(g-r) + \beta_z$, where $\alpha_i=1.49$, $\beta_i= -0.12$, $\alpha_z=1.65$, and $\beta_z=-0.133$.  We assume $r-w=0.1$, and we ignore differences between the Pan-STARRS1 filter system and that used by \citet{Bernardinelli.2022}.

 Finally, we propagate the set of control orbits to the times of each exposure to determine the objects' sky-plane locations and apparent magnitudes as observed from Pan-STARRS1 (Appendix~\ref{APP:Ephemeris}).  Given its 7~sq. deg field of view, a Pan-STARRS1 exposure will typically encompass 1-2 objects from the sample.  We record any whose sky-plane position lands within 1.6~deg of the center of an exposure.  We refer to these as ``candidate'' detections.

\begin{table}[h!]\centering
\begin{tabular}{c| c c c c c | c } 
filter  & g & r & i & z & w & total \\ [0.5ex] 
\hline
$n_{exp}$ & 62,098 & 106,270 & 226,912 & 68,064 & 245,210 & 708,554 \\
\hline
$n_{fov}$ & 120,366 & 207,377 & 444,111  & 132,297 & 479,384 & 1,383,535\\
$n_{det}$ & 31,248 & 66,494 & 135,485 & 31,515 & 222,960 & 487,702 \\
\end{tabular}
\caption{Summary of exposures and synthetic data. $n_{exp}$ is the number of exposures for each filter. $n_{fov}$ is the number of candidate detections, from a control population of 10,000 isotropic orbits, that fall within the field of view of an exposure.  $n_{det}$ is the number of detections that remain after the focal plane filtering stage.}
\label{TAB:candidate_detections}
\end{table}

The number of times that a given object is detected in Pan-STARRS1 exposures will be the primary determinant of whether that object is ``discovered'' by our search algorithms.  In turn, the number of detections of that object is controlled by its the location on the sky and its apparent magnitude.    Before filtering detections that are too faint to detect or that landed on insensitive regions of the focal plane, we can estimate an upper limit on the number of detections we can expect by assuming that all candidate detections that fall within 1.6 deg of the center of an exposure pass the subsequent filtering processes.
Table~\ref{TAB:candidate_detections} shows the outcomes for the 10,000 objects in the control sample for all the exposures in our survey, as a function of filter.  Each exposure has $\sim2$ candidate detections.

\begin{figure}[!htp]
\centering
\includegraphics[trim = 0mm 0mm 0mm 0mm, clip, angle=0, width=\columnwidth]{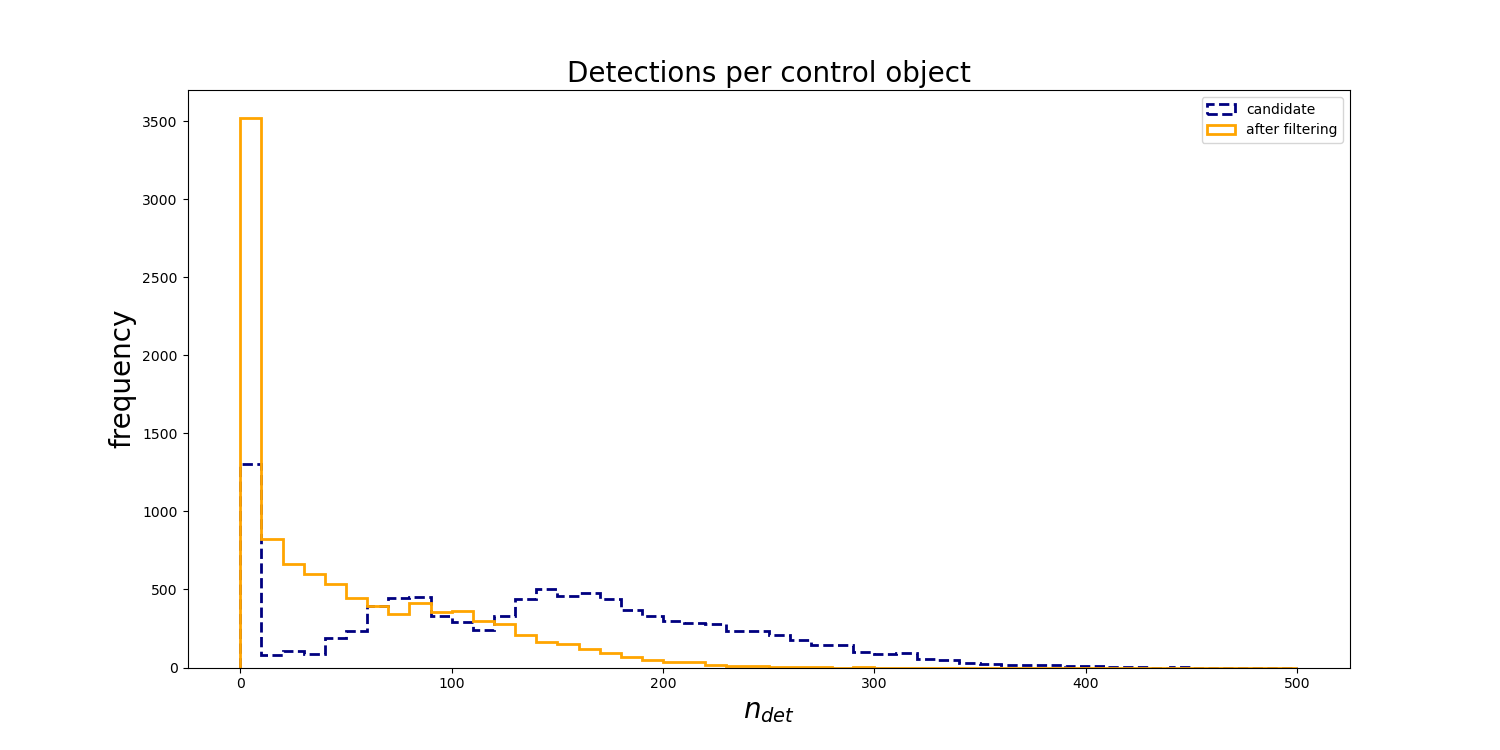}
\caption{%
The number of detections per control object. The dashed blue line shows this distribution of candidate detections, those that fall within the field of view of an exposure.  Of the 10,000 control objects, 8747 land within the circular boundary at least one exposure.  Of those that hit at least one exposure, the median number of candidate detections is 154.  The solid orange line shows the distribution of detections that remain after the filtering steps.  As expected, the distribution is shifted to smaller numbers of detections, and the number of objects that go undetected is significantly larger.  Of those that are detected in at least one exposure, the median number of detecions is 43.  The shapes and medians of the histograms depend upon the details of the control population, but the post-filtering histogram gives us a rough estimate of the expected number of detections for real objects.
}
\label{FIG:ndet}
\end{figure}

Figure~\ref{FIG:ndet} shows the number of candidate detections per control object.  Of the 10,000 objects in the isotropically distributed control sample, 8747 of them fall within the field of view of at least one exposure in the survey.  This highlights the breadth of sky coverage of the Pan-STARRS1 surveys.  Each object has a median of 154 candidate detections, thus there is the potential to discover even those that are bright enough to detect only 10-20\% of the time.

At this stage we need to account for the difference between the model position and magnitude of each candidate detection and what would actually be observed, given the astrometric and photometric uncertainties. 
For each candidate detection, we start with its model RA/Dec and magnitude.
We use the {\it global} World Coordinate System (WCS) to determine which chip in the focal plane the candidate detection might land in, if any. 
Given the chip, its zero point, the predicted instrumental magnitude of the object, and the empirical distribution for the astrometric and photometric uncertainties (see Appendix~\ref{APP:ASTUNC}), we sample the position and magnitude from those distributions, leading to a right ascension and declination (RA/Dec) that is slightly modified from the model values.
 We then convert the slightly modified RA/Dec to an x, y pixel position on the chip, using the {\it local} WCS for that chip.
We also record the estimated astrometric and photometric uncertainties.  Candidate detections that fall outside the limits of the detectors are eliminated at this stage.

\section{Filtering Detections}
\label{SECN:Filtering}

The vast majority of Pan-STARRS1 detections do not correspond to moving objects.  Typically they represent real astrophysical sources (stars and galaxies), features around saturated stars, background fluctuations from the sky, streaks from artificial satellites, or camera artifacts.  A key step to speeding up our search for moving objects is eliminating as many of the detections as possible that are unlikely to be moving sources while minimizing the loss of those that could correspond to our targets of interest.  

\begin{figure}[!htp]
\centering
\includegraphics[trim = 0mm 0mm 0mm 0mm, clip, angle=0, width=0.45\columnwidth]{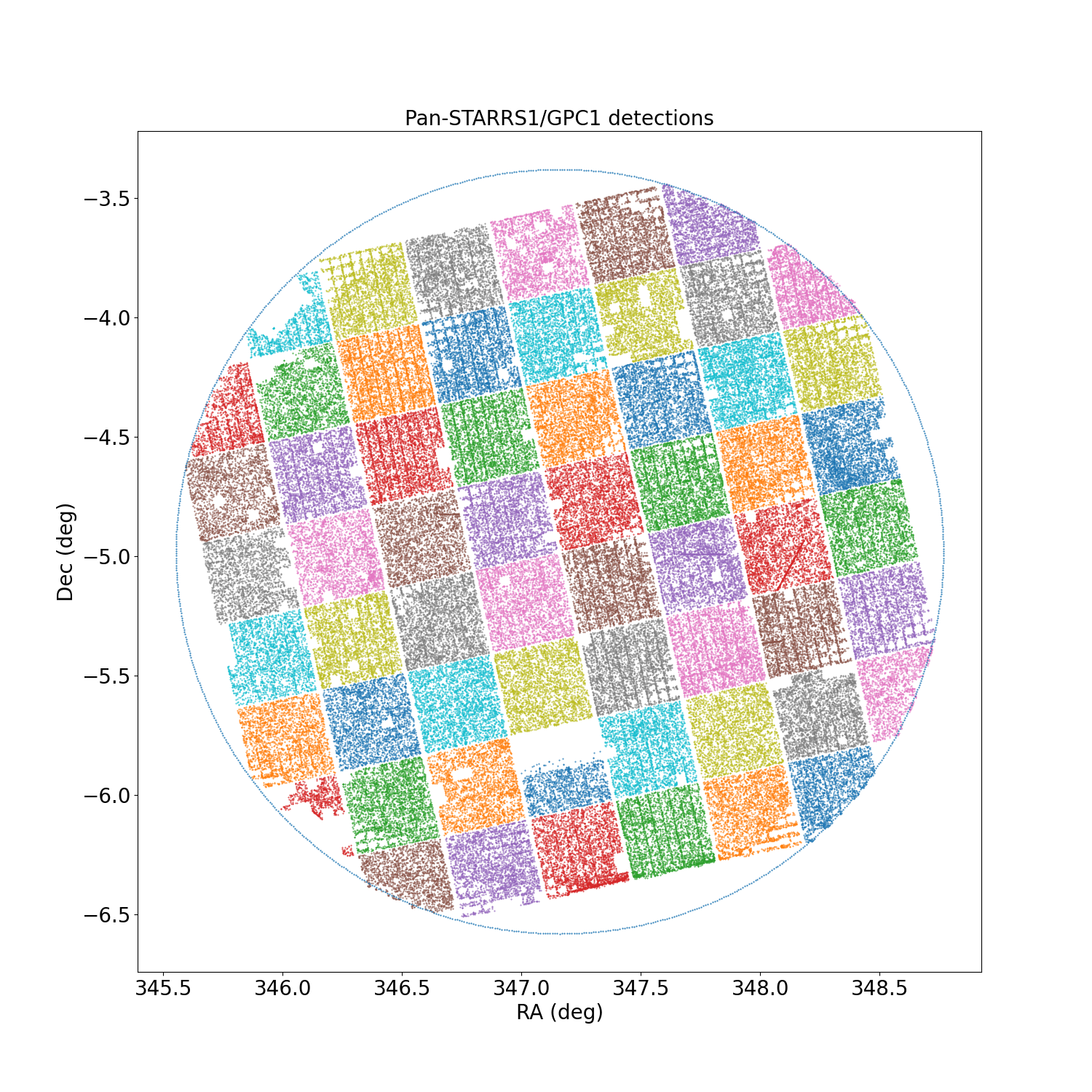}
\includegraphics[trim = 0mm 0mm 0mm 0mm, clip, angle=0, width=0.45\columnwidth]{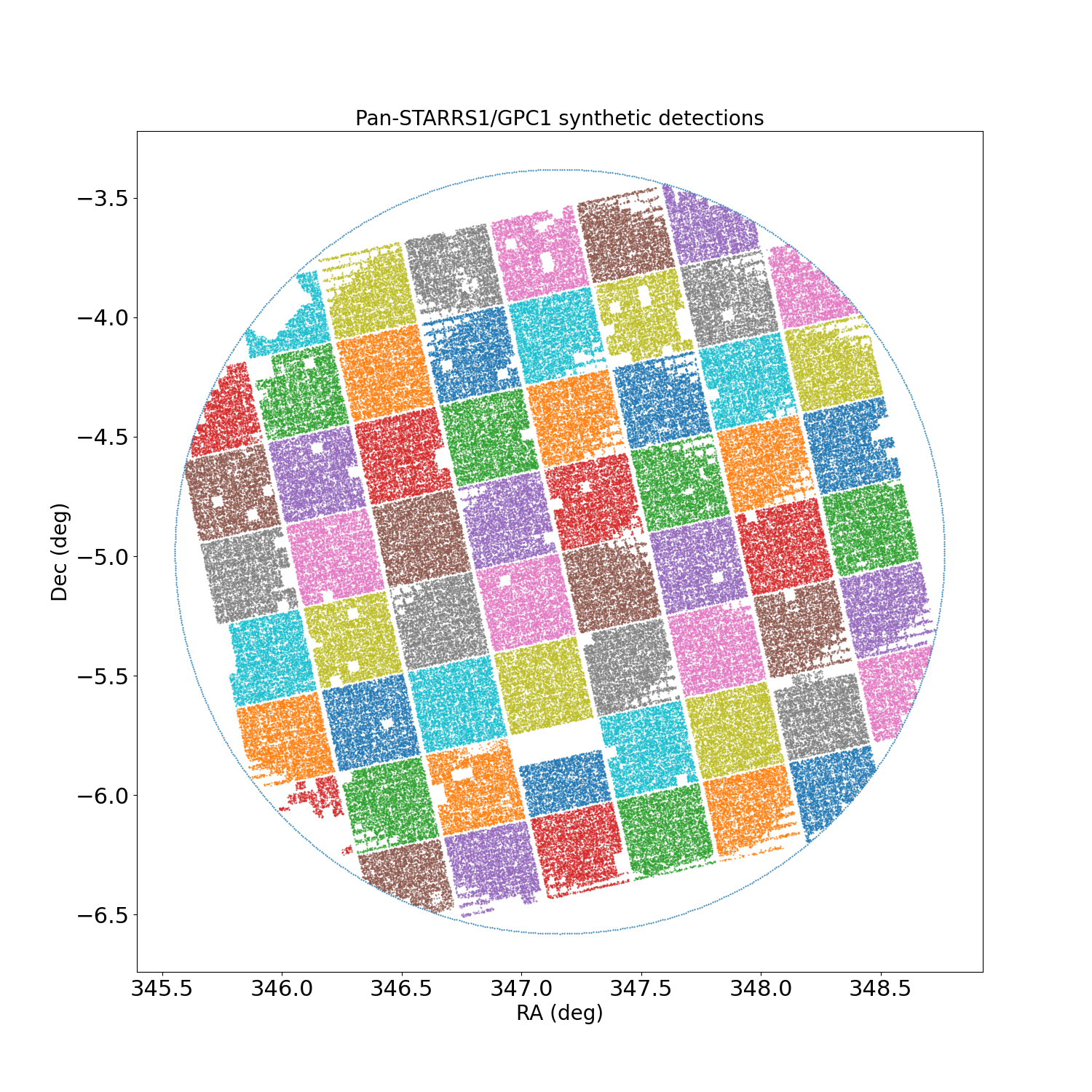}
\caption{%
Left: The sky-plane locations of all detections reported in a single GPC1 exposure.   The colors highlight the sixty individual detectors.  
The 167,750 detections are distributed in rough proportion to the number of live pixels in each of the detectors.  The $8\times8$ grid of cells within each detector is evident.  Some cells are inactive.  There is an excess of detections along the edges of each cell.  The corners of the focal plane are rounded are due to a combination of vignetting and the deliberate placement of poor quality detectors near the edges of the field of view.  The dotted blue circle, with an angular radius of $1.6\deg$, indicates the approximate GPC1 field of view.
Right: The result of placing $4\times 10^5$ synthetic detections uniformly within a circle with an angular radius of $1.6\deg$, centered on the exposure shown in Figure~\ref{FIG:GPC1}.  The detections that fell in the inter- and intra-chip gaps, on the ``hot edges'' or other regions that generate spurious detections, or on pixels that are flagged as bad in the static masks have been removed, following the filtering process described in Section~\ref{SECN:Filtering}. Roughly 68\% of input detections remain after filtering.
}
\label{FIG:GPC1}
\end{figure}

Figure~\ref{FIG:GPC1} highlights the challenge of doing this correctly.  It shows the sky-plane locations of the detections in a single GPC1 exposure and illustrates a number of features of the telescope, camera, and detectors.  There are 60 detectors.  The field is not aligned with RA and Dec; the Pan-STARRS1 telescope has an altitude-azimuth mount and a field rotator.  Vignetting limits the field of view in the corners of the focal plane.  The detections are distributed more or less uniformly, in proportion to the amount of live area in each detector.  The individual readout cells, which form an $8\times8$ grid within each detector, can be clearly seen.  Some cells are not functioning, as is evident from a lack of detections in those areas.  An excess of detections on the perimeter of each cell is also clear.

The detection catalogs generated by the IPP include many attributes that could be used to distinguish real from spurious sources or to otherwise aid the identification of moving objects.  However, we only consider those features that we can reliably and accurately generate for synthetic sources.  For example, the IPP generates a host of flags to indicate the quality of a detection, but we have no way of generating such flags without injecting sources into the images and re-running the IPP.  Thus, we do not use that information in our search.

We eliminate three major categories of detections: (1) those that appear in particularly noisy regions in the focal plane and are likely to be spurious; (2) those that appear near known {\it saturated} astrophysical sources and are also likely to be spurious; and (3) those that appear at the locations of known {\it unsaturated} astrophysical sources and are likely to be real but stationary.

The filtering process only requires knowledge of the sky position of the detection or its corresponding position within the focal plane of a specific exposure.  Thus, the same process can be applied to real and synthetic sources.

\subsection{Noisy Regions of the Focal Plane}
Here we identify regions of the focal plane where a significant excess of detections is generated.  These regions are common to all exposures.  The first of these are the ``hot edges'', the waffle pattern seen in Figure~\ref{FIG:GPC1}.  Each detector is composed of an 8x8 grid cells (590 pixels x 598 pixels of active detector area) that can be read out independently. 
Each detection comes with an x, y pixel position for its particular detector.  There are gaps of 18 pixels between cells in the $x$ direction and 12 pixels in the $y$ direction~\citep{Tonry.2012a}.  
A large number of detections occur within a few pixels of the edges of the cells.  Nearly all of these are spurious.
To identify them empirically, we compute $\bar{x} = x\,\%\,608$ and $\bar{y} = y\,\%\,610$, where $\%$ is the modulo operator.
We then retain only those detections with $12<\bar{x}<581$ and $7<\bar{y}<587$.  This is the same for each cell in each of the detectors. This process excludes about 6\% of the active pixel area but eliminates a few tens of percent of likely false detections.

In addition to the ``hot edges'', there are other fixed regions that generate an overabundance of detections but that vary from detector to detector.  To identify these we subdivided each cell into a 100x100 grid of subcells.  We then examine the catalogs of thousands of exposures spanning a broad range of nights, accumulating a count of the number of detections that appear in each of the subcells.  Those subcells with a detection that is $10\,\sigma$ above the mean are recorded for future reference.  Any detection that lands in one of those subcells is eliminated.

Finally, we eliminate detections that are near pixels that would be masked out by the IPP during image processing.  There are sets of pixel mask images that are used during image processing.    There are ``static'' masks that depend only upon the detectors.  As the name suggests, these remain the same from exposure to exposure.  In addition, there are ``dynamic'' masks that depend upon the sources in a specific image (bright stars, internal reflections, etc).    Each has a mask image for each of the 64 cells in each of the 60 detectors.  Many detector pixels underlie each detection.  In principle the central pixel could be bad while the detection is overall reliable because it is supported by other good pixels.  However, the distribution of good and bad pixels is contiguous enough that the central pixel adequately represents the reliability of the detection.  We use only the static masks and exclude any detection whose central pixel has any mask value that indicates any problem.  Later we assess the impact of ignoring the dynamic masks.

\subsection{Detections Near Saturated Stars}

The IPP reports an excess of detections near bright stars.  These result from excess charge from the star bleeding into adjacent pixels or from the diffraction spikes radiating from the star.  As in \citet{Brown.2024}, we developed an empirical expression for the ``saturation radius,'' the size of the region surrounding a bright star from which detections will be eliminated.  (The saturation radius is $R_{sat} = 0\farcs286\, S^{0.3}$, 
where $S$ is the instrumental flux of the bright star.) This depends primarily upon the instrumental magnitude of the star; brighter stars have larger diffraction spikes and can saturate a larger number of pixels.  The instrumental magnitude in turn depends upon the color of the star, the filter of the exposure, the duration of the exposure, and its measured zero point.  We use the ATLAS catalog to obtain the locations, proper motions, magnitudes, and colors of the bright stars~\citep{Tonry.2018}.  We assume that the $w$-band magnitudes of the bright stars are the same as their $r$-band magnitudes.  We use the predicted locations of known asteroids to verify that detections of real sources are rare within the saturation radius.  That is, saturation both generates spurious detections and suppresses real detections, as expected.

\subsection{Stationary Sources}

The vast majority of astrophysical objects detected by \PS~are stationary on the time scale of the survey.  These objects are observed at nearly the same location each time they are detected.  Separating the detections of stationary objects from those of moving solar system bodies is a key step in our pipeline.

Subtracting pairs of images, or a high signal-to-noise ratio (SNR) reference image from the individual images, is one approach to removing stationary sources~\citep{Alard.2000,Holman.2004,Napier.2024a}.  This approach can be highly effective, but it also has drawbacks.  Image subtraction often generates a host of artifacts, and it can degrade the limiting magnitude, particularly for background-limited sources~\citep{Denneau.2013}.  Furthermore, unless the reference image is composed of images from epochs distinct from those being searched, some of the signal will remain in the template and thus be subtracted from the individual exposures, reducing the SNR of the detection.

We adopt a simpler approach that has been used in a number of previous searches~\citep{Trujillo.2001,Petit.2006,Bannister.2016}.  That is, we reject any detections that coincide with a known stationary source.  Most other TNO surveys are designed to support this kind of rejection using only the exposures taken on any individual night.  The exposures are usually taken near opposition, to maximize the apparent rates of motion.  They typically span $\sim 2-4$~hours within a night.  Given the apparent rates of motion of outer solar system objects, and the accuracy of the astrometry, this allows a clear distinction between stationary and moving objects, even for objects that are hundreds of au away~\citep{Trujillo.2014}.  

However, because the Pan-STARRS1 images of a given region of sky on a typical night span from tens of minutes to an hour, we cannot reliably distinguish stationary objects from moving objects using only the data within a night, without an unacceptably high false positive rate.  Instead, we use a catalog of sources that appear to be stationary over time spans of years.   Any catalog of stationary sources that is nearly complete to the depth of the individual exposures would work, assuming the positions of the sources in the catalog are accurately known.
After some experimentation, we opted to develop our own stationary catalog from the detections in the individual, direct exposures.

We then eliminate any detections that lie within 1\arcsec~of a source in the stationary catalog, leaving only those detections that are spurious, moving, or were not bright enough to appear in the catalog.  This has a negligible impact on the effective survey area.
There are $\sim1.5\times 10^5$ detections in a typical exposure, covering 7~sq.~deg.   The solid angle within 1\arcsec~of those sources is $0.036$~sq.~deg, or 0.5\% of the solid angle of the exposure.   The probability of a chance alignment between a moving object and a stationary source is low, far smaller than the probability of falling in the gaps between or within the detectors.  For a typical Pan-STARRS1 exposure, roughly 80-95\% of those detections will coincide with a known stationary source. Thus, we can reduce the number of detections to be searched by an order of magnitude at the cost of $\sim0.5\%$ of the detections of moving objects.

\begin{figure}[!htp]
\centering
\includegraphics[trim = 0mm 0mm 0mm 0mm, clip, angle=0, width=\columnwidth]{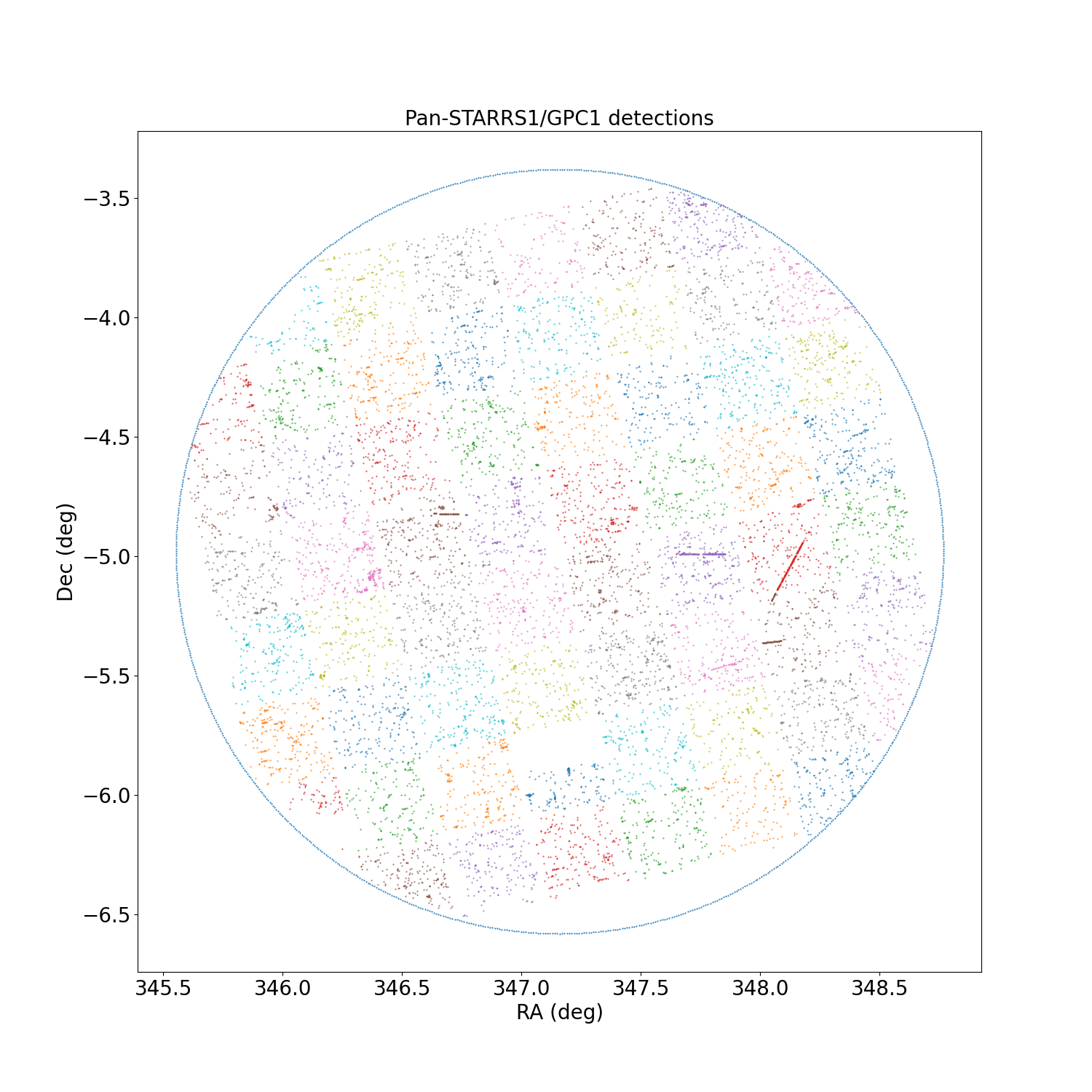}
\caption{%
Of the 167,750 detections shown in Figure~\ref{FIG:GPC1}, only 10,050 remain after the removal of those that 
(1) fall in a noisy region of the focal plane, (2) are close to a bright star, or (3) coincide with a known stationary source. For this exposure, $\sim94\%$ of the detections are removed by one or more of these criteria.   
}
\label{FIG:GPC1_remainder}
\end{figure}
%

The right panel of Figure~\ref{FIG:GPC1} demonstrates the result of injecting $4\times 10^5$ synthetic detections uniformly within a circle with an angular radius of $1.6^\circ$ centered on the exposure shown in left panel and then filtering those according to the steps just described, with the exception of the final detection efficiency step.  Of the input detections, $68\%$~remain.  That is, the fill factor of the GPC1 within an angular radius of $1.6^\circ$ is 0.68.   The filtering process reproduces the details of the focal plane and the detectors with high fidelity.

In Figure~\ref{FIG:GPC1_remainder}, we show that sky-plane locations of the detections that remain after eliminating those that (1) fall in a noisy region of the focal plane, (2) are close to a bright star, or (3) coincide with a known stationary source.  Of the 167,750 detections in the corresponding exposure, only 10,050 remain.  
The long streaks in the remaining detections are likely the result of satellite trails.  Other linear features coincide with the edges of detectors cells.  Some of those could be removed with more aggressive masking.  And some of the detections are sky fluctuations.  We leave removing more of these like likely spurious detections to future work.

\subsection{Detection Probability}

Up to this point the real and synthetic detections have been treated identically, but the real detections have gone through one additional step, implicitly. 
The photons associated with the real detection represent a sample from a Poisson distribution.  For each real detection recorded by the IPP, the total counts exceeded a $5\,\sigma$ detection threshold.  This process must be simulated for the synthetic detections.  For each candidate synthetic detection, we retrieve the parameters for the detection efficiency model for the exposure and chip in question (Appendix~\ref{APP:DET_EFF}).  The model represents the probability that a source of a given instrumental magnitude would exceed the detection threshold for that exposure and detector.  For each candidate detection, we draw a sample uniformly between 0 and 1.  Only those candidate detections for which the uniform sample is less than the model probability are retained.

We record information for the real and synthetic detections that remain after filtering.
This includes the RA, Dec, magnitude, SNR, x \& y pixel coordinates, and the uncertainties associated with these values.  As noted earlier, we only use properties that can be generated for synthetic detections.

The process of filtering stationary and spurious detections depends only upon the RA/Dec and pixel positions, and each detection is treated independently.  Because we will likely use other control populations in future searches, we remove the stationary and spurious detections from the files of real sources separately from the synthetic detections, to avoid unnecessarily reprocessing the real detections.  To carry out a search with another control population, we only need to remove the stationary and spurious detections from that control population and then combine it with the real detections that remain after filtering.

\section{Search Pipeline}
\label{SECN:PIPELINE}


After we have generated synthetic detections from a control population, combined those with the real detections, and filtered out a significant fraction that are unlikely to be our quarry, the typical exposure contains $\sim1000-2000$ real detections and $~1-2$ synthetic detections.   And now we are ready to search for moving objects among the remaining detections.

In an abstract sense, we are trying to separate the detections into many disjoint categories.  We have already identified many of the detections as likely to be {\it stationary} or {\it spurious}.  Now we are attempting to further divide the remaining detections into several small groups, each of which represents a distinct solar system object, and one large group that is everything else.

What distinguishes a set of detections that represent the same solar system object is that there is an orbit that passes through those detections.  That is, there is a dynamical state (position and velocity vectors) at some reference time that, when numerically integrated to the times of the observations, will appear near the observed sky positions when viewed from the position of the observatory.  Furthermore, there should be some absolute magnitude and associated light curve properties that similarly match the observed brightness of the object.  If an orbital and photometric solution exist, there is in fact an ensemble of orbits and associated photometric properties that pass near the observations.  

Our task is to find those solutions.  In principle, one can exhaustively test a multi-dimensional grid of orbital solutions and identify those that pass near many data points or maximize the accumulated signal in a sequence of images~\citep{Bernstein.2004}.  There are bases that simplify the process of enumerating a complete set of orbits~\citep{Bernstein.2000,Holman.2019,Napier.2024a}.  We plan to explore that approach in a companion paper (part 2).  In this paper, we use a more conventional method, but one that is sensitive to very slow-moving objects. We will do this in a sequence of steps, gradually constructing sets of detections that correspond to the same moving object.

\subsection{Making Tracklets in Spherical Barycentric Coordinates}

A moving source observed over a short period of time will have a sequence of detections near the same location, consistent with a linear, constant rate.  We call these ``tracklets,'' following the terminology of \citet{Kubica.2007}.   In this step, we further filter the detections by identifying all physically plausible tracklets and discarding sources that are not part of at least one tracklet. 

For the purposes of our search, the detections comprising a tracklet must be from the same night's observations.  Pan-STARRS1 exposures are typically taken in overlapping patches, cycling through a sequence of pointings two to four times.  For each night, we separate the exposures into contiguous groups.  Because these groups are disjoint both on the sky and in time, the detections for a tracklet will fall entirely within a single group.  Therefore, we can process them in parallel.

Linking the detections of solar system objects in topocentric coordinates suffers from the age-old problem that the apparent motion of solar system objects on the sky executes a looping pattern of prograde and retrograde motion.   However, if we were to observe from the Sun or solar system barycenter, the apparent motion would follow a great circle.   Transforming the observations to a heliocentric or barycentric vantage requires  the heliocentric, barycentric, or geocentric distance, which is unknown {\it a priori}.  However, asserting a distance, or a set of distances, and proceeding is a highly effective means of linking detections~\citep{Holman.2018a}.   We use this approach for both identifying tracklets within a night and for linking tracklets between nights.

We define $\mathbf{r_{obs}}$ as the barycentric position vector of the observatory at the time of the observation and $\mathbf{\hat{\rho}}$ as the unit vector from the observatory in the observed direction of the object.  Both $\mathbf{r_{obs}}$ and $\mathbf{\hat{\rho}}$ are known precisely.  For an asserted barycentric distance, $r$, we can write 
\begin{equation}
r^2 = \rho^2 + r_{obs}^2 - 2\rho r_{obs} \cos \phi , 
\label{EQN:R}
\end{equation}
where $\cos\phi= - \hat{\rho} \cdot \hat{r}_{obs}$.
The solution to Eqn.~\ref{EQN:R} for the topocentric distance, $\rho$, permits zero, one, or two real solutions for $\rho$:
\begin{equation}
\rho = r_{obs}\cos \phi \pm \sqrt{r^2 - r_{obs}^2\sin^2\phi}.
\label{EQN:RHO}
\end{equation}
\noindent{}%
(For distances beyond Earth, there can be at most one solution.)  We ignore solutions for which $\rho < 0 $, which implies the observer is looking in the opposite direction (i.e. through the Earth).  
For each real, positive solution for $\rho$ it is then possible to determine the corresponding barycentric position vector $\mathbf{r}$ of the observed body.  
Rather than using the full position vector, we use its unit vector $\mathbf{\hat{r}}$.  
This is equivalent to using barycentric right ascension and declination or ecliptic longitude and latitude, depending upon which coordinate system is adopted.

In addition to following a great circle, the angular rate of motion for a {\it bound} orbit is limited to being less than that of parabolic motion for an object at pericenter at the assumed distance.  The maximum angular separation of any two detections that are associated with the same solar system object is $ d\theta = \dot{\theta}_{max} \Delta t $, where $\dot{\theta}_{max}=\sqrt{2GM_b/r^3}$, $GM_b$ is the gravitational constant of the solar system, and $\Delta t$ is the time elapsed between the detections.  (In practice, we add in quadrature a small value to $ d\theta$ to accommodate the astrometric uncertainty of the points.)

The algorithm for finding all possible tracklets in a  group of exposures is as follows.  
Given the set of real and synthetic detections in the group, we find the time span $\Delta t$ from the first exposure of the patch to the last.  
For each assumed distance we transform the topocentric detection and load the resulting barycentric unit vectors for the detections in the group into a 3-D KD-tree.
Then for each detection in the group, we use the KD-tree to find all neighbors within an angular separation $d\theta$.  We eliminate any neighbors that have a time that is less than or equal to that of the query detection or that have a difference in magnitude greater than 1.  The remaining detections form tracklets of length two with the query detection.  The first condition ensures that the pairs of detections are from distinct exposures and that we have only one instance of each pair.
Next, we attempt to iteratively extend each pair in this neighborhood by searching for matching detections at the locations predicted by linear, constant rate motion.  We eliminate any tracklets whose residuals exceed 0.4 arcsec, that have a maximum magnitude range greater than 1, or that is a subset of a longer valid tracklet.
At the end of this process we have for each group a set of tracklets for each assumed distance. 

\subsection{Linking Tracklets}

Given the reduced set of detections that comprise the tracklets, we must associate those between nights.  This is known as ``linking tracklets.''   We process the data in chunks, by region on the sky, using the HEALPix tesselation ($n_{side}=32$), as in \citet{Holman.2018a} and \citet{Kurlander.2025}.  

Each HEALPix tile is defined by the position of its center.  We define an inertial coordinate system with the origin at the solar system barycenter, and the $z$ axis pointing toward the center of the HEALPix tile (see Appendix~\ref{SECN:Algorithms}), defining a local tangent plane.  The $x$ axis points in the direction of increasing ecliptic longitude.  And the $y$ axis points toward increasing ecliptic latitude.    We follow the approach and notation of \citet{Bernstein.2000}, restated here for completeness.    

In the local tangent plane, the angular coordinates of the observations are given by
\begin{equation}
\theta_x(t) = { {x(t^\prime) - x_E(t)} \over {z(t^\prime) - z_E(t)}}, \, \theta_y(t)  =  { {y(t^\prime) - y_E(t)} \over {z(t^\prime) -
z_E(t)}, }
\label{exact}
\end{equation}
where $t^\prime=t-\Delta t$, and $\Delta t$ is the light travel time from the object to the observer.  The observatory position ($x_E$, $y_E$, $z_E$) is known precisely.
The trajectory of the target body can be written as 
\begin{equation}
{\bf x}(t) = {\bf x}_0 + {\bf \dot x}_0 t + {\bf g}(t), 
\end{equation}
where ${\bf g}(t)$ is the deviation from linear motion due to the gravitational acceleration toward the barycenter.  It is given by integrating
 $\ddot {\bf g}(t)  \approx  -GM_T \frac{{\bf x}(t)}{|{\bf x}(t)|^3}$, assuming ${\bf g} = 0$ and $\dot {\bf g}= 0$ at $t=0$.
$GM_T$ is the total gravitational constant of the Sun and planets.  Note that ${\bf g}(t)$ is small for $t\ll~T_{orb}$, where $T_{orb}$ is the orbital period of the object.
We ignore the perturbations of the individual planets and massive asteroids; they are negligible for outer solar system bodies on the time scale of years.

The \citet{Bernstein.2000} parameterization, based on the components of ${\bf x}_0$ and ${\bf \dot x}_0$ at the reference time, 
\begin{equation}
\label{definitions}
\begin{array}{lllll}  
\alpha \equiv {x_0}/{z_0} &,& \beta \equiv {y_0}/{z_0} &,& 
	\gamma \equiv 1/{z_0} \\ 
\dot\alpha \equiv {\dot x_0}/{z_0}&,& \dot\beta \equiv {\dot y_0}/{z_0} &,& 
	\dot\gamma \equiv {\dot z_0}/{z_0},
\end{array}
\end{equation}
supports a simple description of the motion of outer solar objects.
In terms of these parameters,  the observations $\theta_x$ and $\theta_y$ are well modeled by
\begin{equation}
    \begin{split}
        \theta_x(t) & =  { { \alpha + \dot\alpha t^\prime + \gamma g_x(t^\prime) - \gamma x_E(t) }
        \over { 1 + \dot\gamma t^\prime + \gamma g_z(t^\prime) - \gamma z_E(t) } }
        \\
        \theta_y(t) & = { { \beta + \dot\beta t^\prime + \gamma g_y(t^\prime) - \gamma y_E(t) }
        \over { 1 + \dot\gamma t^\prime + \gamma g_z(t^\prime) - \gamma z_E(t) } },
    \end{split}
\label{posexpression}
\end{equation}
where $t^\prime \approx t - \Delta t$ and $\Delta t = \frac{1}{c \gamma}$.
We can rearrange equations~\ref{posexpression} to yield simple expressions for the linear motion of the object:
\begin{equation}
\begin{array}{lccl}  
        \phi_x(t) = { { \alpha + \dot\alpha t^\prime }} &=& 
                    \theta_x(t) f(t^\prime) + \gamma x_E(t)\\
        \phi_y(t) = { { \beta + \dot\beta t^\prime }}  &=& 
                    \theta_y(t) f(t^\prime) + \gamma y_E(t),
\label{rearrange}
\end{array}
\end{equation}
where $f = 1 + \dot\gamma t^\prime + \gamma g_z(t^\prime) - \gamma z_E(t)$.  We ignore transverse components of the gravitational perturbation, $g_x(t^\prime)$ and $g_y(t^\prime)$, as they are much smaller than the radial component $g_z(t^\prime)$.  We also make the approximation that $g_z(t^\prime) = -{1\over2}GM_T \gamma^2 {t^\prime}^2$, which is equivalent to assuming that the motion of a distant solar system body is ballistic with constant acceleration toward the barycenter.  

The above equations provide the essence of the search technique we use in this investigation. If we assume values of $\gamma$ and $\dot\gamma$ and ignore $g_x$ and $g_y$, all quantities on the right hand sides of equations~\ref{rearrange} are known or observed. The quantities on the left-hand sides are simply linear, constant-rate motion, because the calculations on the right-hand sides account for the parallax from the observatory and the small scale variations due to changes in the distance to the object. We adopt a series of $\gamma$-$\dot\gamma$ pairs, transform the observations to $\phi_x$ and $\phi_y$ for each of those parameters pairs, and look for linear motion within the transformed observations. We set $\dot\gamma=0$ and choose uniformly distributed values of $\gamma$, corresponding to distances of 80.0, 84.2, 88.9, 94.1, 100.0, 106.7, 114.3, 123.1, 133.3,  145.5, 160.0, 177.8, 200.0, 228.6, 266.7, 320.0, 400.0, 533.3, 800.0, and 1600.0 au.

One final complication is that there is too much data to do a full-sky search in one pass. To get around this limitation, we process the data in chunks, by exploiting the property that for bound orbits, we have $\dot\alpha\leq n_{para}$ and $\dot\beta \leq n_{para}$, where $n_{para} = \sqrt{2 GM_t \gamma^3}$.  The maximum angular displacement from the location at the mid-time of the survey (JD 2456635.5 TDB) is given by $\theta = {T\over 2} n_{para}, $ where $T$ is time span of the data ($\sim9$~years).  For each search region, we include the data within the angular radius $\theta$. This ensures that all plausibly associated detections are processed together.

Then for each chunk of data and for each $\gamma$-$\dot\gamma$ pair, we proceed as follows:
\begin{enumerate}
    \item Transform the detections according to equations~\ref{rearrange}.
    \item Identify all tracklets. 
    \item Choose a pair of tracklets that can be plausibly fit to a bound orbit. See if the orbit can be extended to another tracklet, and remain bound.  Repeat this to exhaustion.
    \item Repeat step 3 for all pairs of tracklets.
\end{enumerate}

Since our chunks on the sky overlap and the objects move, and since some objects may be linked at multiple $\gamma$-$\dot\gamma$ pairs, the above procedure sometimes results in repeated linkages. We merge such linkages by clustering them using a metric that we call a similarity score; linkages that share many detections will have a high similarity score, and thus be merged. 

Finally, our linking procedure sometimes allows false positive detections into a candidate. We eliminate such detections by running a tune-up routine in which we iteratively eliminate outliers.  

Although we explored a range of distances from 80~au to 1600~au, the algorithm has some sensitivity to objects outside of that distance range (see below).  That is because if three tracklets, in transformed coordinates, lie close enough to a line in the tangent plane (step 3 above), the full orbit from fitting the three tracklets is used to search for additional tracklets, which are then also included in the fit.  However, the search is much less complete outside of the search range.

Our linking procedure completed in several days on 100-200 cores on Harvard's Cannon Cluster.  It is difficult to more accurately assess the total computation required, due to the queuing policies on the cluster, but this estimate is good to a small factor.

\section{Qualitative Analysis}
\label{SECN:QA}

\begin{figure}[h]
    \centering
    \includegraphics[width=0.8\linewidth]{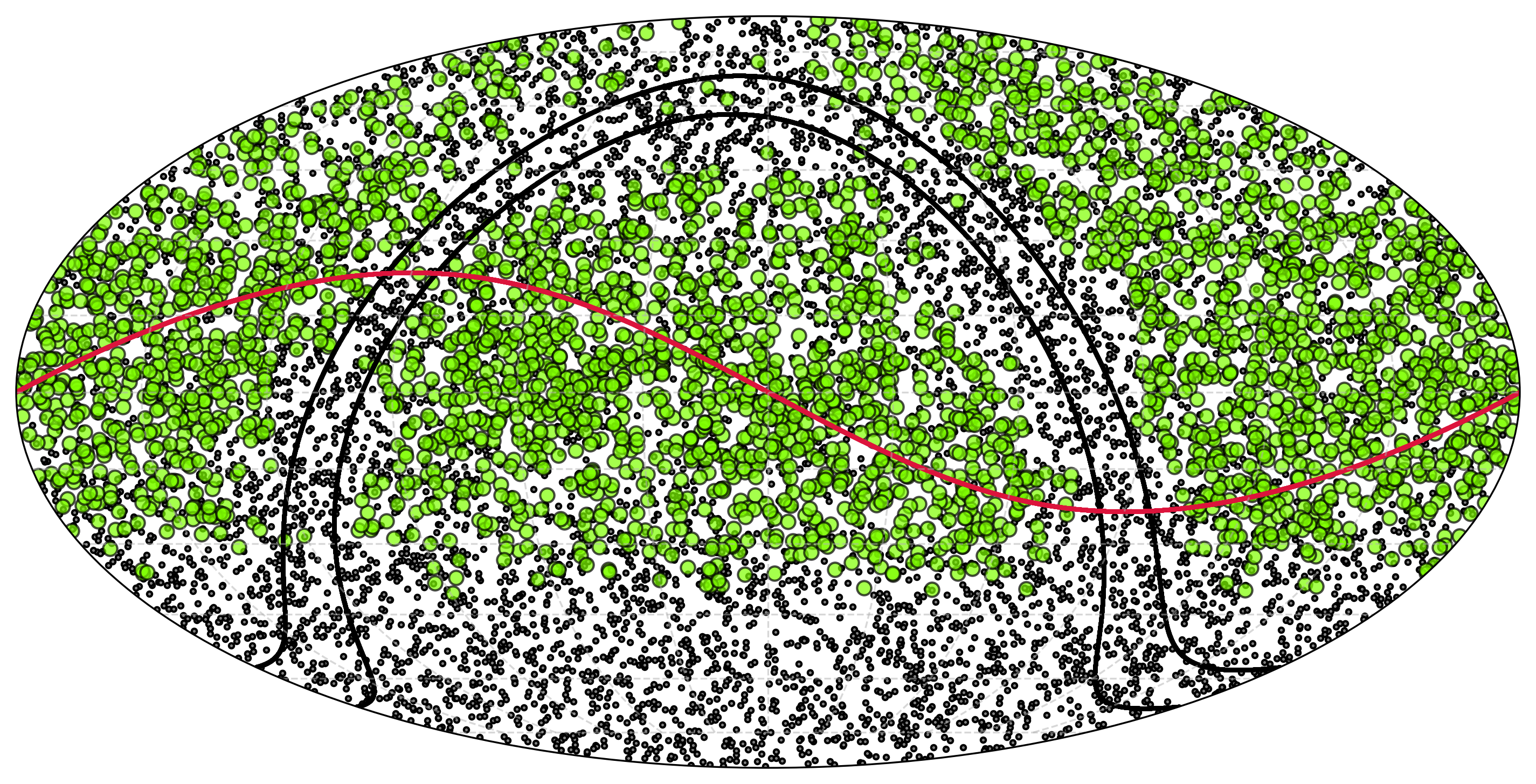}
    \caption{Sky positions of our synthetic objects relative to F51 on 8 December 2013 (approximately the midpoint of our data). The orientation is the same as in Figure~\ref{FIG:COVERAGE}. The green and black points represent objects that our simulator predicts would be recovered and missed, respectively. Note the strong spatial dependence introduced by the galactic plane.}
    \label{fig:fakes}
\end{figure}

With our linkages in hand, we decided to discard any that were composed of fewer than 6 tracklets, as a lower threshold resulted in a flood of false positive linkages. A post-hoc analysis of our implanted synthetic sources justifies this choice, as the contamination rate of the sources constituting their linkages was of order $0.05\%$. After this cut, our final list of linkages consisted of $3807$ candidate objects, $3115$ of which turned out to be implanted synthetic sources, leaving us with $692$ candidates composed of purely real detections, $109$ of which are not yet listed in the Minor Planet Center's database. It is clear in Figure \ref{fig:fakes} that objects near the galactic plane were strongly biased against discovery, mostly because there were not enough images in those regions for us to securely detect them using our pipeline. 

Of the $692$ candidate objects that were not implanted synthetic sources, we found $23$ objects with $H_w < 4$. All of these objects were found to be previously known large TNOs and dwarf-planets.  Of the $31$ presently known outer solar system objects with $H_w < 4$, the ones that we did not find were too bright and thus saturated (Pluto), too faint to detect, or near the galactic plane regions where our survey has limited data.  It is encouraging that we were able to find such a large fraction of the known massive minor planets, but nonetheless disappointing that no new such object was discovered. Among the objects detected in our search were Eris (96 au), Gonggong (87 au), Sedna (86 au), and 532037 (80 au), demonstrating that our survey is sensitive to distant objects. However, we did not find any new objects beyond 60 au. Although our search did not turn up any new distant objects, our inclusion of implanted synthetic sources still allows us to run the population studies on large, distant objects (see Sections \ref{sec:simulator} and \ref{sec:p9}).

\section{Survey Simulator}
\label{sec:simulator}
One of the primary goals of this work is to enable hypothesis testing of solar system populations via survey simulation. The basic idea is extremely simple: posit a set of orbits, and determine which objects would have been detected by the survey. Doing this correctly involves a detailed understanding of a survey's detection efficiency, as well as a reliable model for determining whether the detections of an object would have been correctly linked. We already have a detailed understanding of the survey's detection efficiency (as we demonstrated earlier in this paper), and we are able to predict with high fidelity which images any given object would have been detected in. The final step, then, is to determine whether we would have successfully linked an object's detections. 

\subsection{Linking Model}
Rather than re-running our linking algorithm for each population we want to simulate, we can use the results of our implanted synthetic sources to develop a model that predicts whether a source would have been linked. One can imagine any number of ways to build such a model. For example, an extremely simple heuristic would be to simply declare that we would have been able to link any object that produced at least 6 tracklets over the course of the survey. We could perhaps further refine the model by requiring that the tracklets occurred over some number of oppositions. The process can be arbitrarily complicated, and the correct approach depends strongly on the survey strategy and the linking algorithm (see \citealt{Bernardinelli.2022} for one example of a hand-tuned criterion). Instead of choosing heuristics in an ad-hoc manner, we have chosen to take a machine learning approach, so that we can explore whether there is any complicated dependence on the cadence of the detections.

\begin{figure}[h]
    \centering
    \includegraphics[width=0.5\linewidth]{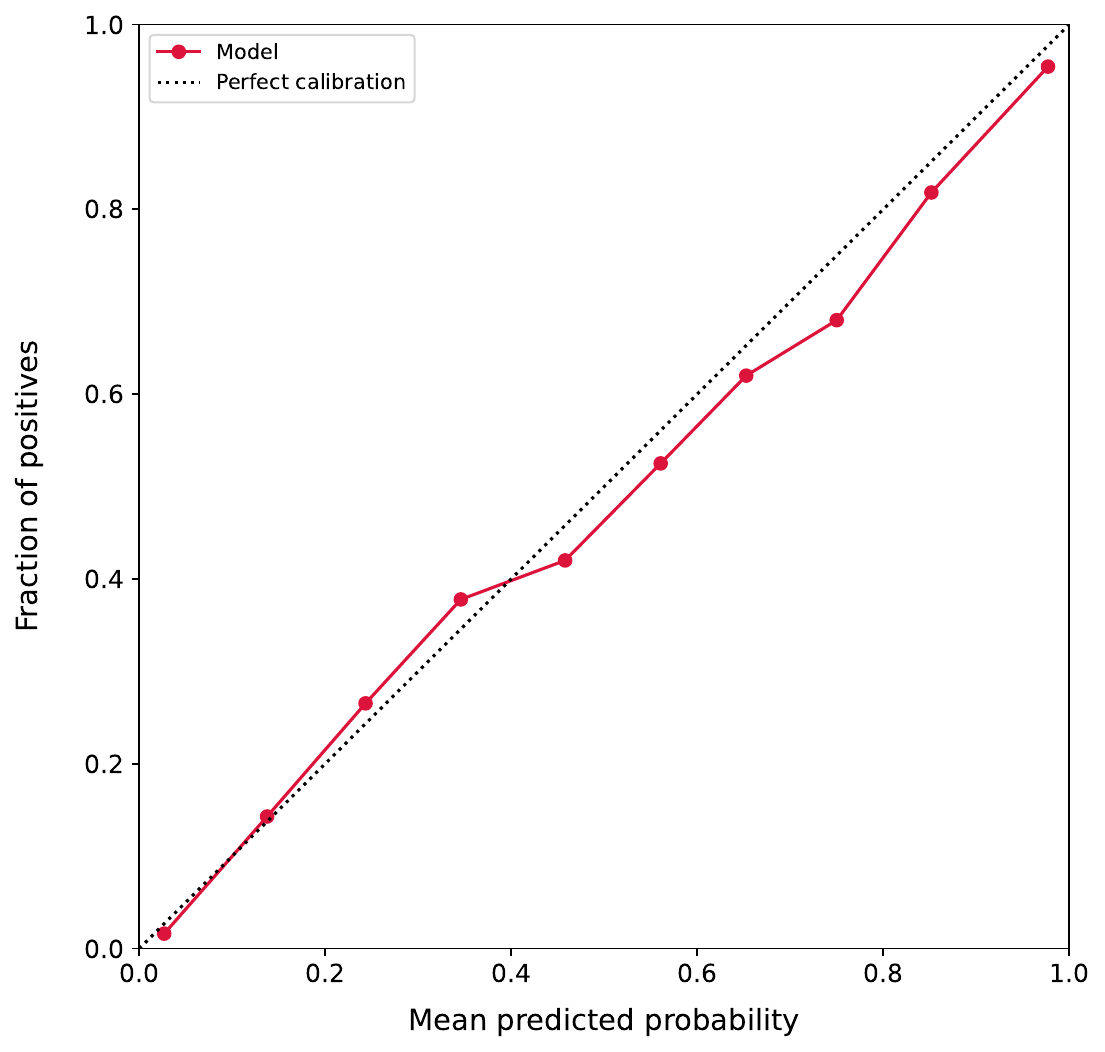}
    \caption{Calibration plot for our logistic regression model. The points on the x axis are binned values of the predicted probability of a vector being linked, while the y axis represents the fraction of the objects in that bin that were actually linked. The model, only deviates from perfect calibration by a few percent, indicating that it is justifiable to treat its predictions as probabilities.}
    \label{fig:calibration}
\end{figure}

As a first attempt at building a model, we used a simple neural net to perform a logistic regression. We encoded each object's detections into a single vector by accumulating the number of detections in bins of 10 days, and labeled the vector with a 1 if the object was successfully linked, and a 0 if it was not. (The choice of 10-day bins is arbitrary, but we found that it worked well.)  We reserved 20\% of our data as a validation sample, and trained our neural net with the remaining 80\%. Using this simple model we were able to predict with 90\% accuracy whether our validation set was linked, and we did not find any obvious biases among the misclassified objects. More importantly, our model is very well-calibrated, meaning that the probability it assigned for any given vector to be linked was commensurate with the fraction of such vectors that were actually linked (see Figure \ref{fig:calibration}). For example, about 20\% of the vectors with scores of 0.2 were actually linked. This property implies that we can reasonably interpret our model's predictions as \textit{probabilities}, thus making it an appropriate tool for Monte Carlo population studies.

Although the logistic regression model happened to work extremely well on our data, we note that this is not always the case. Sometimes linking may depend sensitively on a survey's cadence in ways that are not easily described by using a linear model. To test for this sensitivity in our data, we also trained a model using a recurrent neural net (RNN), and found nearly identical results to our simpler model.\footnote{The inputs to the RNN were slightly different from the inputs for the linear model. Since RNNs can handle variable-length sequences, we did not need to encode the information into a fixed-length vector. Instead we were able to provide the model with lists of the epochs of the detections of an object, which we ``normalized'' by subtracting off the start time from each epoch.} The recurrent neural net has clear advantages over a simple linear model, in that it can model nonlinearity, and can handle variable-length sequences, so that the data does not need to be binned. However, training and evaluating an RNN is much more computationally intensive than a linear model, and the meaning of the model is more opaque. Using an RNN did not provide any improvement over our linear model, so we have chosen to use the linear model for this particular application.

\subsection{Validation with a Second Population of Fakes}
\begin{figure}[h]
    \centering
    \includegraphics[width=0.8\linewidth]{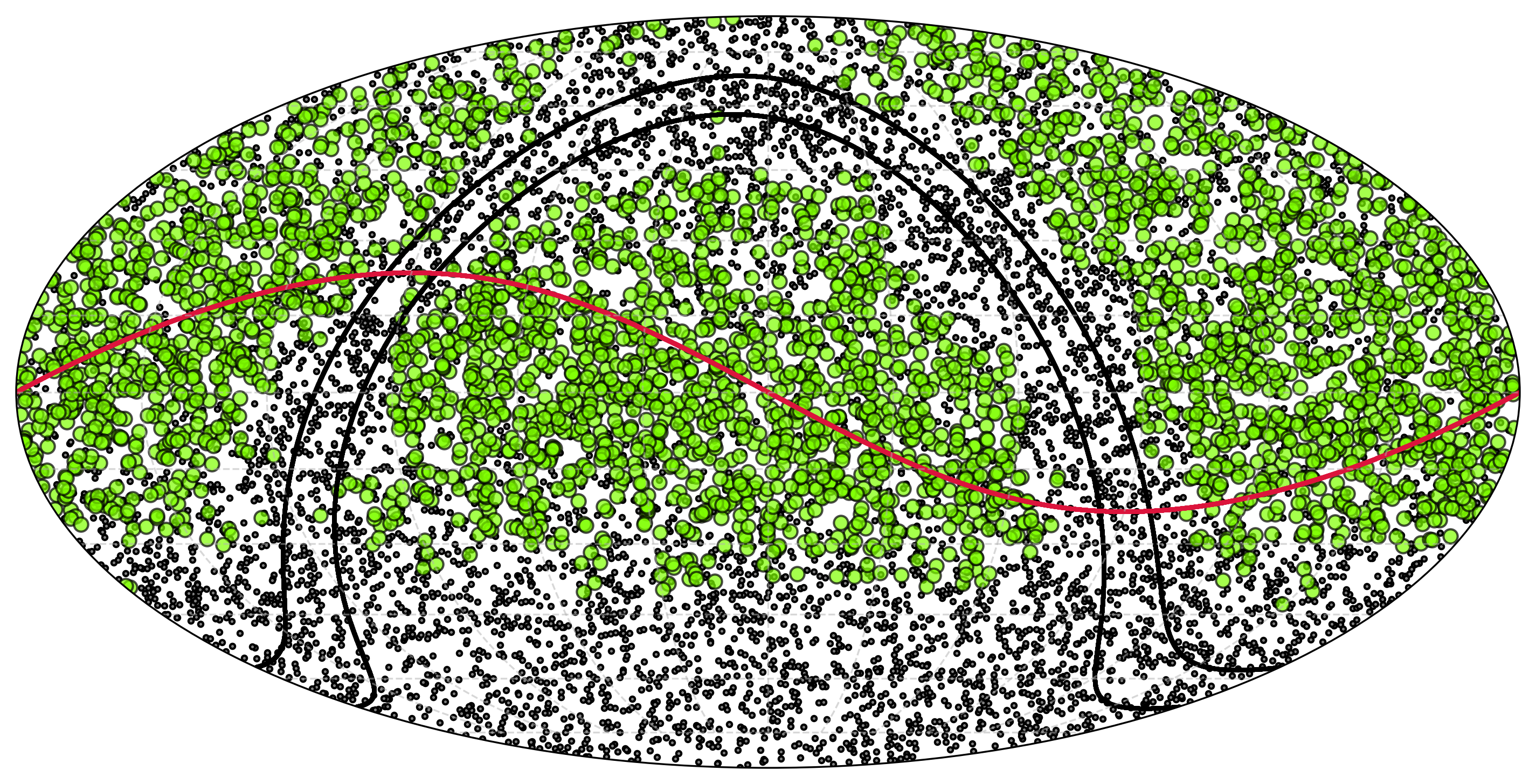}
    \caption{Sky positions of our second set of synthetic objects relative to F51 on 8 December 2013 (approximately the midpoint of our data). The orientation is the same as in Figure~\ref{FIG:COVERAGE}. The green and black points represent objects that our simulator predicts would be recovered and missed, respectively. Note the strong resemblance to Figure \ref{fig:fakes}.}
    \label{fig:validation}
\end{figure}

As a simple first validation test for our survey simulator, we generated a second population of 10,000 synthetic objects using the same process that we used to generate the control population used in our search. This test is not meant to be a statistically rigorous, but rather to provide a second source of validation that our simulator produces realistic results for an input population that it was not explicitly trained on. The sky distribution of the objects that the simulator predicts would be found (see Figure \ref{fig:validation}) closely matches the distribution of the control population shown in Figure \ref{fig:fakes}, indicating that our simulator is producing qualitatively correct results.

\section{Limits on Planet Nine}
\label{sec:p9}

We now use our simulator on the Planet Nine reference population first described by \cite{Brown.2021}, and provided by \cite{Brown.2023}. By running this population through our simulator we can determine how much of the Planet Nine parameter space has been ruled out by non-detection in this analysis.

The reference population is composed of 100,000 synthetic objects that are consistent with the current Planet Nine models, and includes orbital elements, size, albedo, and a $V$-band magnitude. These parameters provide us with almost enough information to run the population through our survey simulator; however, since the the catalog does not specify colors for the objects, we are left to choose them ourselves. We opt to use the TNO colors provided by \cite{Bernardinelli.2022}, and the \cite{Jester.2005} transformations to convert a $V$-band magnitude to the corresponding $wgriz$ colors.

\begin{figure}[H]
    \centering
    \includegraphics[width=0.8\linewidth]{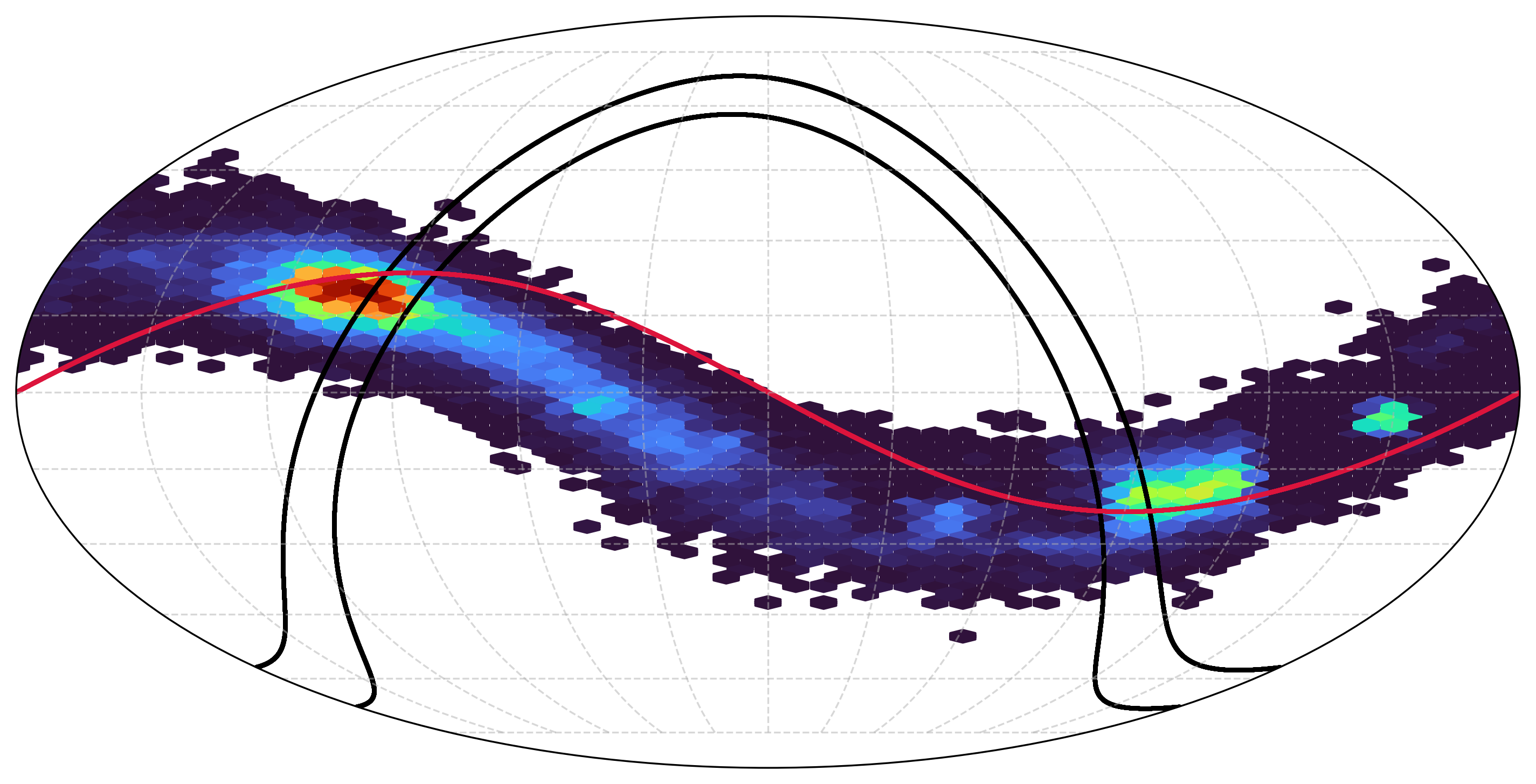}
    \includegraphics[width=0.8\linewidth]{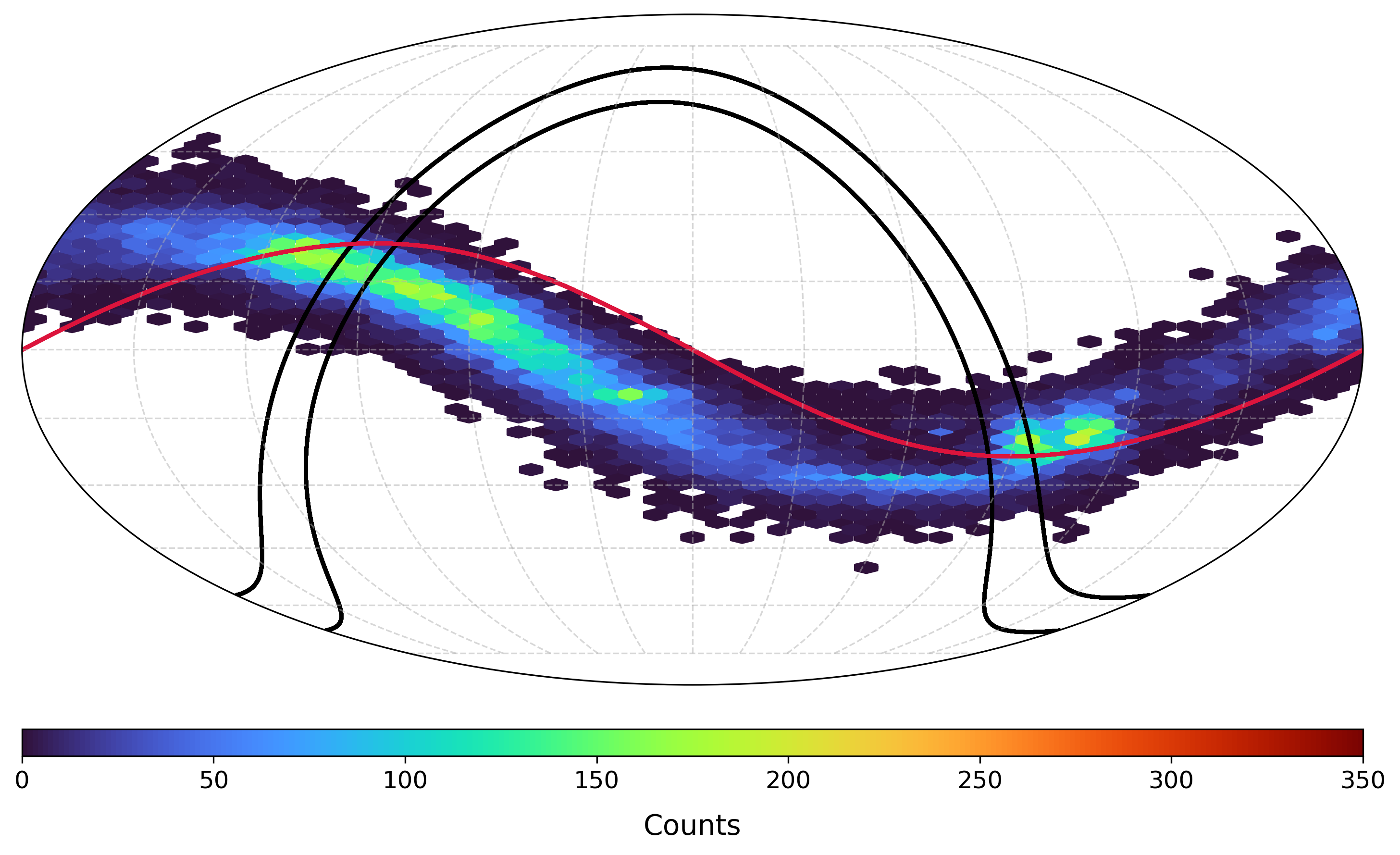}
    \caption{On-sky density of the objects in the Planet Nine reference population that have not yet been ruled out by our survey (top) and by \citet{Brown.2024} (bottom). The orientation of each plot is the same as in Figure~\ref{FIG:COVERAGE}}
    \label{fig:p9-remaining}
\end{figure}

Our survey simulator predicts that we would have discovered 75,769/100,000 of the synthetic Planet Nine objects at $50\%$ confidence. \cite{Brown.2024} analyzed a smaller subset of the Pan-STARRS source catalog we used in this study, and ruled out 68,745/100,000 of the objects in the reference catalog. We show the on-sky density of the objects that have \emph{not} been ruled out by our study in Figure \ref{fig:p9-remaining}. Our results are qualitatively consistent with those of \citet{Brown.2024}.  In principle, we could combine our results with those of \citep{Brown.2024} to place tighter constraints on the existence of Planet Nine.  However, our respective approaches to determining the efficiency, or probability, that an object with a given orbit and magnitude would be discovered are sufficiently different that combining the results would require quite a bit of care.  Furthermore, the two searches used a lot of the same data, so the results would be correlated.  Given those challenges, we defer combining the results to future work.

The "hot spot" on the far right of the top panel of figure~\ref{fig:p9-remaining} is primarily the result of the difference between our algorithm and that of \citet{Brown.2024}. Although our search includes a larger dataset, it was less sensitive in regions where there is limited data.  That is, more detections are required to secure a discovery.  In the galactic plane regions, where little to no w-band exposures were taken, our limits are weaker.  However, we stress that our approach of  implanting synthetic sources into the catalogs, rather than using known asteroids to characterize a detection efficiency, allows us to test the sensitivity of the search in areas where there are few known asteroids.

\section{Conclusions}
\label{SECN:Conclusions}

We have completed a search for very distant solar system objects using data from the Pan-STARRS1 survey using a method that is sensitive to a wide range of distances, as well as rates and directions of motion. Our search resulted in the detection of $692$ solar system objects, $109$ of which are not yet listed in the Minor Planet Center's database. By raw number, our 642 TNO detections makes our search the third most productive Kuiper Belt survey to date, behind the Outer Solar System Origins Survey (838; \citealt{Bannister.2018}) and the Dark Energy Survey (812; \citealt{Bernardinelli.2022}).\footnote{We found 642 objects with $a > 30$ au. We found 554 objects with $q > 30$ au, in case the reader prefers that definition.} Note that we did not explicitly search for objects at barycentric distances closer than 80 au; we will search at closer heliocentric distances in a followup to this work.

We calibrated our search by injecting a control population at the catalog level, being careful to reproduce the detailed characteristics of the camera focal plane. We also developed and employed a novel technique for survey simulation that enables the linking step to be treated probabilistically. By combining the two techniques, we were able to develop a survey simulator that is well suited for statistical population studies.

Although we did not find Planet Nine (or any other planetary objects), we were able to use our survey simulator to show that the remaining parameter space for Planet Nine is highly concentrated in the galactic plane. As a followup to this work, we plan to more thoroughly search the galactic plane region (as well as the rest of the survey) using a trackletless linking algorithm, which we believe will enable us to confidently discover objects using fewer detections.

Finally, we note that our detections included 23 dwarf planets. Unfortunately all of these objects were previously known; however, the fact that we were able to detect all of them in one well-characterized search means that we will be able to set new limits on the total population in future work.

\section*{Acknowledgments}

 MJH gratefully
acknowledges support from the NSF (grant No. AST2206194), the NASA YORPD Program (grant No.
80NSSC22K0239), and the Smithsonian Scholarly Studies Program (2022, 2023).
KJN is supported by Schmidt Sciences, LLC.
MJP acknowledges support from the NASA Minor Planet Center Award 80NSSC22M0024 and
from NASA award 80NSSC20K0641.

The Pan-STARRS1 Surveys (PS1) have been made possible through contributions of the Institute for Astronomy, the University of Hawaii, the Pan-STARRS Project Office, the Max-Planck Society and its participating institutes, the Max Planck Institute for Astronomy, Heidelberg and the Max Planck Institute for Extraterrestrial Physics, Garching, The Johns Hopkins University, Durham University, the University of Edinburgh, Queen's University Belfast, the Harvard-Smithsonian Center for Astrophysics, the Las Cumbres Observatory Global Telescope Network Incorporated, the National Central University of Taiwan, the Space Telescope Science Institute, the National Aeronautics and Space Administration under Grant No. NNX08AR22G issued through the Planetary Science Division of the NASA Science Mission Directorate, the National Science Foundation under Grant No. AST-1238877, the University of Maryland, and Eotvos Lorand University (ELTE).

The computations in this paper were run on the Cannon cluster supported by the FAS Science Division Research Computing Group at Harvard University.


\appendix

\section{Time}
\label{SECN:Time}

It is critical to have accurate times of observation when searching for moving objects and determining their orbits. The IPP records multiple sources of time within the Pan-STARRS1 telescope and camera system.   After cross checking, we conclude that the shutter open time, in UTC, plus half the exposure time is a reliable exposure mid-point.  We convert these time to TDB for subsequent calculations.

\section{HEALPix and KD-trees}
\label{SECN:Algorithms}

We make extensive use of two data structures to organize and efficiently process Pan-STARRS1 data: HEALPix\footnote{http://healpix.sf.net} and KD-trees.

The HEALPix tessellation of the sky \citep{Gorski.2005} divides the unit sphere into equal area, uniformly distributed regions. HEALPix first partitions the unit sphere into twelve equal area, roughly square base pixels.  
Each of these twelve base pixels is then subdivided into an $N_{side}\times N_{side}$ grid of pixels, where $N_{side}$ is a power of two.   
For our purposes, the principal advantage of the HEALPix tiling for our is {\it healpy}, an efficient python library of routines that support transformations and other calculations between coordinates on the unit sphere and HEALPix pixel indices~\citep{Zonca.2019}.
For this paper, we adopt the $N_{\mathrm{side}}=32$ resolution level and orient the HEALPix grid with equatorial coordinates.  This results in 12,288 HEALPix tiles that are $1\fdg83 \times 1\fdg 83$, each centered on a specific RA \& Dec.   With $N_{\mathrm{side}}=32$, a single GPC1 exposure typically overlaps a few HEALPix tiles.

KD-trees are space-partitioning data structures that support efficient searches in  low-dimensional data sets~\citep{Bentley.1975,Kubica.2007}.  In particular, KD-tree radius searches, finding for all points that are within a given ``distance'' of a single query point, scale as $O(\log N)$ operations, where $N$ is the number of data points in the tree.  Thus, a search for the neighbors of all of the data points can be completed in $O(N\log N)$ operations.  For large data sets, this can be orders of magnitude faster than a brute force search, which requires $O(N^2)$ operations. The KD-tree implementation\footnote{https://docs.scipy.org/doc/scipy/reference/generated/scipy.spatial.cKDTree.query.html} that we use includes a ``dual tree'' algorithm that makes searching for the neighbors of all points in the tree even more efficient~\citep{Maneewongvatana.1999}. 

\section{Astrometric Solution}
\label{APP:WCS}

The IPP fits a two-stage WCS for each exposure using 2MASS and Gaia reference stars.  There is a global WCS that transforms a position in an idealized focal plane to RA/Dec coordinates.  For each detector there a secondary set of equations that transform that detector to the idealized focal plane.

The overall WCS solution translates an x, y pixel position on a given detector in a given exposure to RA, Dec.  This is typical for most mosaic CCD cameras.  However, we need the inverse process for our control population.  We start with the candidate RA, Dec positions in a given exposure and need to determine which detector it might correspond to and then the x, y position on that detector.  

\section{Detection Efficiency Model}
\label{APP:DET_EFF}

As noted in section~\ref{SECN:PS1}, the IPP provides detection efficiency information for each exposure. Five hundred synthetic, stationary point sources with the local stellar PSF are injected into each detector. This is done for values of instrumental magnitude that straddle the expected 50\% detection limit. The IPP records how many of those injected sources are recovered as a function of instrumental magnitude.

The sources are first injected across the full extent of each detector.  Then the bad pixel mask is applied. Some of the detections are lost because they are too faint or land in vignetted regions of the focal plane. Some are lost because they fall between cells. Some are masked where the detector quality is poor.  This implies that the maximum detection efficiency can be less than one, even for bright sources.

Most solar system sources are not trailed in the PS1 exposures; they appear stationary within the time span of the exposure. Given that, the point source detection efficiency information is all that is needed in most cases. moving objects to calibrate an analytic model.)
We model the detection efficiency $f$ as
\begin{equation}
    f = f_{max} \left[ 1 - F(X; \nu) \right],
\end{equation}
where $f_{max}$ is the maximum detection efficiency, F is the cumulative probability of Student's $t$-distribution, $\nu$ is the parameter of the $t$-distribution, and $X = N_\sigma-S/N$ measures how close the observed signal-to-noise ratio $S/N$ is to the target significance $N_\sigma$.
The $t$-distribution is symmetric about zero and is similar to a Gaussian but has heavier tails.  The degree to which the distribution is concentrated in the core versus the tails is controlled by the parameter $\nu$.  
The signal-to-noise ratio is given by
\begin{equation}
    S/N = \frac{S}{\sqrt{S + B}},
    \label{EQ:SNR}
\end{equation}
where $S=10^{-0.4 m_I}$, $m_I$ is the instrumental magnitude of the source, and $B$ is the sum of the non-signal components that contribute to the noise.  In practice, $B$ will be primarily composed of sky background and read noise.   Rather than trying to understand the detailed origin of the noise, we are simply fitting an empirical expression that represents the detection efficiency well.

The IPP records in the smf file headers the instrumental magnitude at which the S/N is expected to be $N_\sigma = 5$.   Given that, we can rearrange eq.~\ref{EQ:SNR} and solve for the value of $B$ for the corresponding signal $S$:
\begin{equation}
     B = (S^2 - S N_\sigma^2)/N_\sigma^2.
\end{equation}
This gives us a reference value for the background.
We then fit for three parameters that characterize the deviations from the expected detection efficiency function: the maximum detection efficiency $f_{max}$, a factor by which the estimated background $B$ needs to be increased or decreased, and the $t$-distribution parameter $\nu$.  These three parameters control the height of the detection efficiency curve, the magnitude at which the detection efficiency drops to half of its maximum, and the slope and shape of the curve through its drop in efficiency.  Although other functional forms and parameterizations are valid,  this one is simple, physically motivated, and works particularly well.

We fit and store the detection efficiency parameters separately for each chip in each exposure and store them to use during the injection of synthetic detections.

\section{Astrometric Uncertainty Model}
\label{APP:ASTUNC}

We use the uncertainties in the astrometric positions in two ways.  First, those uncertainties characterize the distributions we will sample to generate synthetic observed astrometric positions from idealized model positions.  Second, the we will use the $\chi^2$ of an orbit fit to assess the likelihood that a collection of observations belongs to the same object. Thus, it is important that we develop a model of the astrometric uncertainties so that we can make realistic simulated data and so that we evaluate our candidate discoveries.  Fortunately, the process of identifying stationary sources and the model of the detection efficiency allow us to develop such a model.

The positions of sources in the stationary catalogs are the average of up to 200 detections and have astrometric uncertainties that are an order of magnitude smaller than those the individual detections. Each detector in each exposure typically has hundreds to thousands of detections that match sources in the stationary catalog.  The differences in the positions of these matched pairs is dominated by the astrometric uncertainty of the individual detections.  Thus, we can use the distribution of the positional differences to model the individual astrometric uncertainties.

We expect the astrometric uncertainty to scale as 
\begin{equation}
    \sigma^2 = \left(\frac{\mathrm{FWHM}}{\mathrm{SNR}}\right)^2 + \sigma_{sys}^2,
\end{equation}
where $\mathrm{FHWM}$ is the full width at half maximum of the PSF, $\mathrm{SNR}$ is the signal-to-noise ratio of the detection, and $\sigma_{sys}$ is minimum astrometric uncertainty due to systematic errors.  The average $\mathrm{FHWM}$ value for each detector is recorded by the IPP.  The $\mathrm{SNR}$ value can be calculated from the magnitude of the detection and the detection efficiency parameters for the detector (see appendix~\ref{APP:DET_EFF}).

We compute the residuals $dx$ and $dy$ in RA and Dec directions, respectively.
We then compute the covariance between the scaled residuals $dx/\sigma$ and $dy/\sigma$ for all detections that lie within 1\arcsec\, of a stationary source.  This scaling separates the portion of the astrometric uncertainty that depends upon the $\mathrm{SNR}$ from the portion that depends upon the atmosphere and telescope motion.  We now have a means of reliably estimating the 2D astrometric uncertainties and a means of generating samples of astrometric positions from a realistic distribution, both as a function of instrumental magnitude.

\section{Observatory Position}
\label{APP:OBS_POS}

It is essential to have accurate observatory locations for computing topocentric positions.  A table of the Earth-fixed geocentric locations of ground-based observatories is maintained by the Minor Planet Center\footnote{https://minorplanetcenter.net/iau/lists/ObsCodesF.html}.  JPL provides a regularly-updated SPICE kernel for the matrix that rotates from the Earth-fixed ITRF93 frame to the J2000 equatorial frame as a function of time.   We use this matrix to convert the Earth-fixed geocentric observatory locations to the J2000 equatorial frame.  To that we add the barycentric location of the Earth's geocenter, in the same equatorial frame, as given by JPL's DE441 ephemeris from its corresponding SPICE kernel~\citep{Park.2021}.  We use the python Spiceypy package for all SPICE kernel calculations~\citep{Annex.2019}.

\begin{figure}[!htp]
\centering
\includegraphics[trim = 0mm 0mm 0mm 0mm, clip, angle=0, width=\columnwidth]{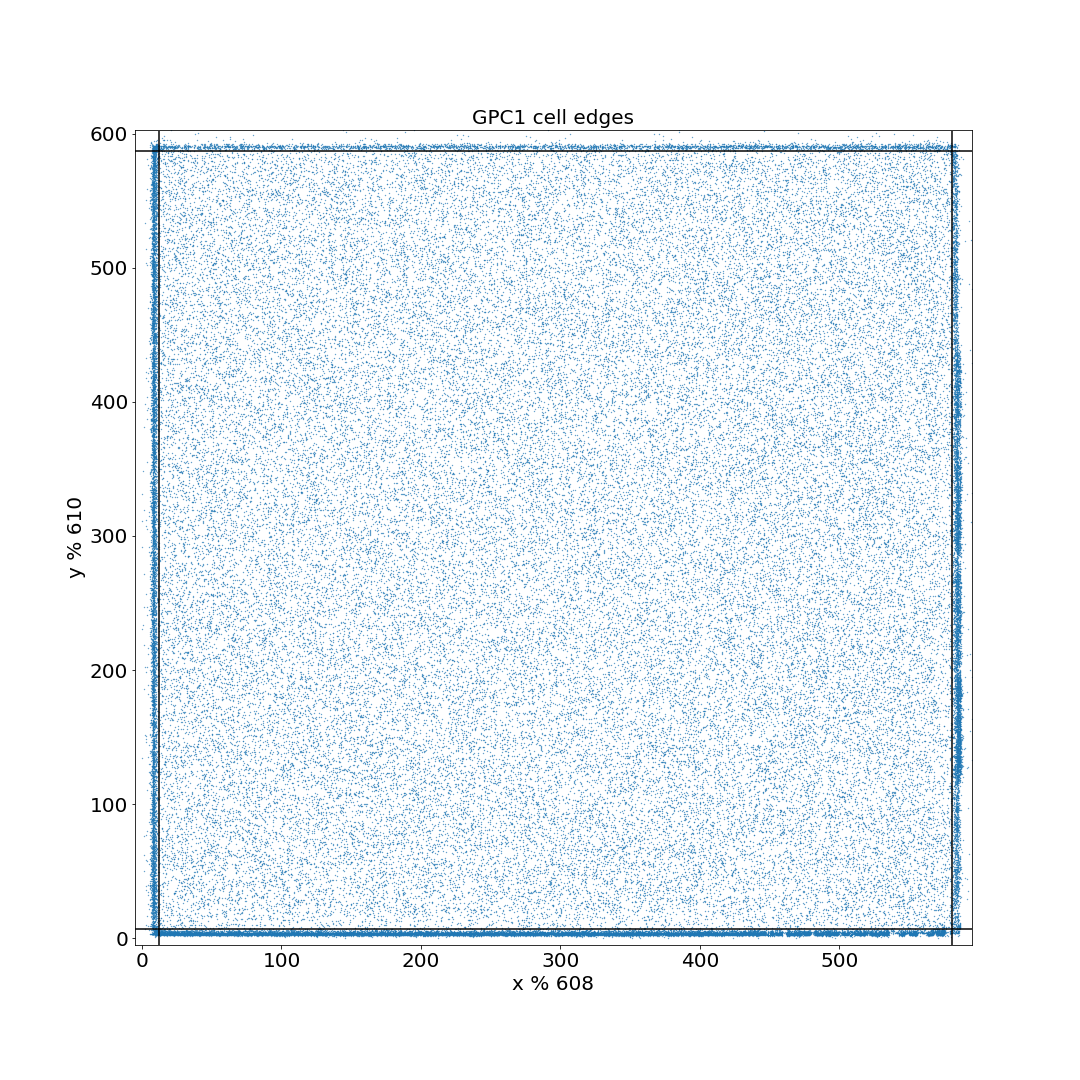}
\caption{%
The detections in all the detectors shown in figure~\ref{FIG:GPC1}, folded to match the repeated cell pattern: $\bar{x} = x\,\%\,608, \bar{y} = y\,\%\,610$.  The overabundance of detections at the borders of the cells is now even more evident.  The horizontal and vertical lines indicate the area trimmed off in our processing.  This eliminates ${\sim}33\%$ of the detections (most of which are false) and ${\sim}6\%$ of the detector area.  
}
\label{FIG:edges}
\end{figure}
%

\section{Ephemeris Generation}
\label{APP:Ephemeris}

The brute force approach to determining which objects appear in which exposure is to numerically calculate the objects' sky-plane positions as seen from the PS1 telescope at the time of each exposure and to then check which of those is angularly close enough to the center of that exposure to plausibly be detected.  However, the apparent angular rates of TNOs observed at opposition, dominated by parallax, is $\sim0.02\deg/day$, and even smaller for more distant objects.  Such objects can take more than a year to traverse the diameter of a GPC1 exposure, which suggests an optimization.  We follow the approach to ephemeris generation develop for \texttt{Sorcha} (Merritt et al. 2025).  (For details see Holman et al. 2025). 

On a nightly basis we compute the light time corrected positions of all the objects as observed from PS1. (See Appendix:\ref{APP:OBS_POS} for details on obtaining the position of the observatory.)
We numerically integrate the orbits as test particles moving in the field of the Sun, planets, Moon, Pluto, and a set of large asteroids using \texttt{ASSIST}~\citep{Holman.2023}.  We then compute HEALPix indices from the observed positions and organize the objects by HEALPix.  Given the RA/Dec position of the center of an exposure we determine the HEALPix tiles that are overlapped by an angular radius of $1.8\deg$.  (This is larger than the  $\sim1.6\deg$ radius of GPC1's field of view to generously allow for the motion of the objects.)  For the just the objects in the set of HEALPix tiles covered by the exposure, we calculate precise positions at the time of the exposure and retain only those that are within its field of view.

\software{ASSIST (Holman et al. 2025), HEALPix (Gorski et al. 2005), healpy (Zonca et al. 2019), Scipy (Jones et al. 2001)}



\bibliographystyle{aasjournal}

\begin{thebibliography}{}
\expandafter\ifx\csname natexlab\endcsname\relax\def\natexlab#1{#1}\fi
\providecommand{\url}[1]{\href{#1}{#1}}
\providecommand{\dodoi}[1]{doi:~\href{http://doi.org/#1}{\nolinkurl{#1}}}
\providecommand{\doeprint}[1]{\href{http://ascl.net/#1}{\nolinkurl{http://ascl.net/#1}}}
\providecommand{\doarXiv}[1]{\href{https://arxiv.org/abs/#1}{\nolinkurl{https://arxiv.org/abs/#1}}}

\bibitem[{{Alard}(2000)}]{Alard.2000}
{Alard}, C. 2000, \aaps, 144, 363, \dodoi{10.1051/aas:2000214}

\bibitem[{{Annex} {et~al.}(2019){Annex}, {Carcich}, {Murakami}, {Kulumani}, {de
  Val-Borro}, {Stefko}, {Diaz del Rio}, \& {Seignovert}}]{Annex.2019}
{Annex}, A., {Carcich}, B., {Murakami}, S.-y., {et~al.} 2019, {SpiceyPy: Python
  wrapper for the NAIF C SPICE Toolkit}, Astrophysics Source Code Library,
  record ascl:1903.016.
\newblock \doeprint{1903.016}

\bibitem[{{Bannister}(2020)}]{Bannister.2020}
{Bannister}, M.~T. 2020, in The Trans-Neptunian Solar System, ed.
  D.~{Prialnik}, M.~A. {Barucci}, \& L.~{Young}, 439--453,
  \dodoi{10.1016/B978-0-12-816490-7.00020-5}

\bibitem[{{Bannister} {et~al.}(2016){Bannister}, {Alexandersen}, {Benecchi},
  {Chen}, {Delsanti}, {Fraser}, {Gladman}, {Granvik}, {Grundy},
  {Guilbert-Lepoutre}, {Gwyn}, {Ip}, {Jakubik}, {Jones}, {Kaib}, {Kavelaars},
  {Lacerda}, {Lawler}, {Lehner}, {Lin}, {Lykawka}, {Marsset}, {Murray-Clay},
  {Noll}, {Parker}, {Petit}, {Pike}, {Rousselot}, {Schwamb}, {Shankman},
  {Veres}, {Vernazza}, {Volk}, {Wang}, \& {Weryk}}]{Bannister.2016}
{Bannister}, M.~T., {Alexandersen}, M., {Benecchi}, S.~D., {et~al.} 2016, \aj,
  152, 212, \dodoi{10.3847/0004-6256/152/6/212}

\bibitem[{{Bannister} {et~al.}(2018){Bannister}, {Gladman}, {Kavelaars},
  {Petit}, {Volk}, {Chen}, {Alexandersen}, {Gwyn}, {Schwamb}, {Ashton},
  {Benecchi}, {Cabral}, {Dawson}, {Delsanti}, {Fraser}, {Granvik},
  {Greenstreet}, {Guilbert-Lepoutre}, {Ip}, {Jakubik}, {Jones}, {Kaib},
  {Lacerda}, {Van Laerhoven}, {Lawler}, {Lehner}, {Lin}, {Lykawka}, {Marsset},
  {Murray-Clay}, {Pike}, {Rousselot}, {Shankman}, {Thirouin}, {Vernazza}, \&
  {Wang}}]{Bannister.2018}
{Bannister}, M.~T., {Gladman}, B.~J., {Kavelaars}, J.~J., {et~al.} 2018, \apjs,
  236, 18, \dodoi{10.3847/1538-4365/aab77a}

\bibitem[{{Batygin} \& {Brown}(2016)}]{Batygin.2016}
{Batygin}, K., \& {Brown}, M.~E. 2016, \aj, 151, 22,
  \dodoi{10.3847/0004-6256/151/2/22}

\bibitem[{Bentley(1975)}]{Bentley.1975}
Bentley, J.~L. 1975, Commun. ACM, 18, 509, \dodoi{10.1145/361002.361007}

\bibitem[{{Bernardinelli} {et~al.}(2020){Bernardinelli}, {Bernstein}, {Sako},
  {Liu}, {Saunders}, {Khain}, {Lin}, {Gerdes}, {Brout}, {Adams}, {Belyakov},
  {Somasundaram}, {Sharma}, {Locke}, {Franson}, {Becker}, {Napier},
  {Markwardt}, {Annis}, {Abbott}, {Avila}, {Brooks}, {Burke}, {Carnero Rosell},
  {Carrasco Kind}, {Castander}, {da Costa}, {De Vicente}, {Desai}, {Diehl},
  {Doel}, {Everett}, {Flaugher}, {Garc{\'\i}a-Bellido}, {Gruen}, {Gruendl},
  {Gschwend}, {Gutierrez}, {Hollowood}, {James}, {Johnson}, {Johnson},
  {Krause}, {Kuropatkin}, {Maia}, {March}, {Miquel}, {Paz-Chinch{\'o}n},
  {Plazas}, {Romer}, {Rykoff}, {S{\'a}nchez}, {Sanchez}, {Scarpine}, {Serrano},
  {Sevilla-Noarbe}, {Smith}, {Sobreira}, {Suchyta}, {Swanson}, {Tarle},
  {Walker}, {Wester}, {Zhang}, \& {DES Collaboration}}]{Bernardinelli.2020}
{Bernardinelli}, P.~H., {Bernstein}, G.~M., {Sako}, M., {et~al.} 2020, \apjs,
  247, 32, \dodoi{10.3847/1538-4365/ab6bd8}

\bibitem[{{Bernardinelli} {et~al.}(2022){Bernardinelli}, {Bernstein}, {Sako},
  {Yanny}, {Aguena}, {Allam}, {Andrade-Oliveira}, {Bertin}, {Brooks},
  {Buckley-Geer}, {Burke}, {Rosell}, {Carrasco Kind}, {Carretero}, {Conselice},
  {Costanzi}, {da Costa}, {De Vicente}, {Desai}, {Diehl}, {Dietrich}, {Doel},
  {Eckert}, {Everett}, {Ferrero}, {Flaugher}, {Fosalba}, {Frieman},
  {Garc{\'\i}a-Bellido}, {Gerdes}, {Gruen}, {Gruendl}, {Gschwend}, {Hinton},
  {Hollowood}, {Honscheid}, {James}, {Kent}, {Kuehn}, {Kuropatkin}, {Lahav},
  {Maia}, {March}, {Menanteau}, {Miquel}, {Morgan}, {Myles}, {Ogando},
  {Palmese}, {Paz-Chinch{\'o}n}, {Pieres}, {Malag{\'o}n}, {Romer}, {Roodman},
  {Sanchez}, {Scarpine}, {Schubnell}, {Serrano}, {Sevilla-Noarbe}, {Smith},
  {Soares-Santos}, {Suchyta}, {Swanson}, {Tarle}, {To}, {Varga}, \&
  {Walker}}]{Bernardinelli.2022}
---. 2022, \apjs, 258, 41, \dodoi{10.3847/1538-4365/ac3914}

\bibitem[{{Bernstein} \& {Khushalani}(2000)}]{Bernstein.2000}
{Bernstein}, G., \& {Khushalani}, B. 2000, \aj, 120, 3323,
  \dodoi{10.1086/316868}

\bibitem[{{Bernstein} {et~al.}(2004){Bernstein}, {Trilling}, {Allen}, {Brown},
  {Holman}, \& {Malhotra}}]{Bernstein.2004}
{Bernstein}, G.~M., {Trilling}, D.~E., {Allen}, R.~L., {et~al.} 2004, \aj, 128,
  1364, \dodoi{10.1086/422919}

\bibitem[{{Bowell} {et~al.}(1989){Bowell}, {Hapke}, {Domingue}, {Lumme},
  {Peltoniemi}, \& {Harris}}]{Bowell.1989}
{Bowell}, E., {Hapke}, B., {Domingue}, D., {et~al.} 1989, in Asteroids II, ed.
  R.~P. {Binzel}, T.~{Gehrels}, \& M.~S. {Matthews}, 524--556

\bibitem[{Brown(2023)}]{Brown.2023}
Brown, M. 2023, Planet Nine reference population, Version 3.0,  CaltechDATA,
  \dodoi{10.22002/8fjad-x7y61}

\bibitem[{{Brown} \& {Batygin}(2016)}]{Brown.2016}
{Brown}, M.~E., \& {Batygin}, K. 2016, ArXiv e-prints.
\newblock \doarXiv{1603.05712}

\bibitem[{{Brown} \& {Batygin}(2021)}]{Brown.2021}
---. 2021, \aj, 162, 219, \dodoi{10.3847/1538-3881/ac2056}

\bibitem[{{Brown} {et~al.}(2024){Brown}, {Holman}, \& {Batygin}}]{Brown.2024}
{Brown}, M.~E., {Holman}, M.~J., \& {Batygin}, K. 2024, arXiv e-prints,
  arXiv:2401.17977, \dodoi{10.48550/arXiv.2401.17977}

\bibitem[{{Brown} {et~al.}(2004){Brown}, {Trujillo}, \&
  {Rabinowitz}}]{Brown.2004}
{Brown}, M.~E., {Trujillo}, C., \& {Rabinowitz}, D. 2004, \apj, 617, 645,
  \dodoi{10.1086/422095}

\bibitem[{{Brown} {et~al.}(2005){Brown}, {Trujillo}, \&
  {Rabinowitz}}]{Brown.2005}
{Brown}, M.~E., {Trujillo}, C.~A., \& {Rabinowitz}, D.~L. 2005, \apjl, 635,
  L97, \dodoi{10.1086/499336}

\bibitem[{{Chambers} {et~al.}(2016){Chambers}, {Magnier}, {Metcalfe},
  {Flewelling}, {Huber}, {Waters}, {Denneau}, {Draper}, {Farrow}, {Finkbeiner},
  {Holmberg}, {Koppenhoefer}, {Price}, {Rest}, {Saglia}, {Schlafly}, {Smartt},
  {Sweeney}, {Wainscoat}, {Burgett}, {Chastel}, {Grav}, {Heasley}, {Hodapp},
  {Jedicke}, {Kaiser}, {Kudritzki}, {Luppino}, {Lupton}, {Monet}, {Morgan},
  {Onaka}, {Shiao}, {Stubbs}, {Tonry}, {White}, {Ba{\~n}ados}, {Bell},
  {Bender}, {Bernard}, {Boegner}, {Boffi}, {Botticella}, {Calamida},
  {Casertano}, {Chen}, {Chen}, {Cole}, {Deacon}, {Frenk}, {Fitzsimmons},
  {Gezari}, {Gibbs}, {Goessl}, {Goggia}, {Gourgue}, {Goldman}, {Grant},
  {Grebel}, {Hambly}, {Hasinger}, {Heavens}, {Heckman}, {Henderson}, {Henning},
  {Holman}, {Hopp}, {Ip}, {Isani}, {Jackson}, {Keyes}, {Koekemoer}, {Kotak},
  {Le}, {Liska}, {Long}, {Lucey}, {Liu}, {Martin}, {Masci}, {McLean}, {Mindel},
  {Misra}, {Morganson}, {Murphy}, {Obaika}, {Narayan}, {Nieto-Santisteban},
  {Norberg}, {Peacock}, {Pier}, {Postman}, {Primak}, {Rae}, {Rai}, {Riess},
  {Riffeser}, {Rix}, {R{\"o}ser}, {Russel}, {Rutz}, {Schilbach}, {Schultz},
  {Scolnic}, {Strolger}, {Szalay}, {Seitz}, {Small}, {Smith}, {Soderblom},
  {Taylor}, {Thomson}, {Taylor}, {Thakar}, {Thiel}, {Thilker}, {Unger},
  {Urata}, {Valenti}, {Wagner}, {Walder}, {Walter}, {Watters}, {Werner},
  {Wood-Vasey}, \& {Wyse}}]{Chambers.2016}
{Chambers}, K.~C., {Magnier}, E.~A., {Metcalfe}, N., {et~al.} 2016, arXiv
  e-prints, arXiv:1612.05560.
\newblock \doarXiv{1612.05560}

\bibitem[{{Chen} {et~al.}(2016){Chen}, {Lin}, {Holman}, {Payne}, {Fraser},
  {Lacerda}, {Ip}, {Chen}, {Kudritzki}, {Jedicke}, {Wainscoat}, {Tonry},
  {Magnier}, {Waters}, {Kaiser}, {Wang}, \& {Lehner}}]{Chen.2016}
{Chen}, Y.-T., {Lin}, H.~W., {Holman}, M.~J., {et~al.} 2016, \apjl, 827, L24,
  \dodoi{10.3847/2041-8205/827/2/L24}

\bibitem[{{Chiang} {et~al.}(2007){Chiang}, {Lithwick}, {Murray-Clay}, {Buie},
  {Grundy}, \& {Holman}}]{Chiang.2007}
{Chiang}, E., {Lithwick}, Y., {Murray-Clay}, R., {et~al.} 2007, Protostars and
  Planets V, 895

\bibitem[{{Denneau} {et~al.}(2013){Denneau}, {Jedicke}, {Grav}, {Granvik},
  {Kubica}, {Milani}, {Vere{\v s}}, {Wainscoat}, {Chang}, {Pierfederici},
  {Kaiser}, {Chambers}, {Heasley}, {Magnier}, {Price}, {Myers}, {Kleyna},
  {Hsieh}, {Farnocchia}, {Waters}, {Sweeney}, {Green}, {Bolin}, {Burgett},
  {Morgan}, {Tonry}, {Hodapp}, {Chastel}, {Chesley}, {Fitzsimmons}, {Holman},
  {Spahr}, {Tholen}, {Williams}, {Abe}, {Armstrong}, {Bressi}, {Holmes},
  {Lister}, {McMillan}, {Micheli}, {Ryan}, {Ryan}, \& {Scotti}}]{Denneau.2013}
{Denneau}, L., {Jedicke}, R., {Grav}, T., {et~al.} 2013, \pasp, 125, 357,
  \dodoi{10.1086/670337}

\bibitem[{{Finkbeiner} {et~al.}(2016){Finkbeiner}, {Schlafly}, {Schlegel},
  {Padmanabhan}, {Juri{\'c}}, {Burgett}, {Chambers}, {Denneau}, {Draper},
  {Flewelling}, {Hodapp}, {Kaiser}, {Magnier}, {Metcalfe}, {Morgan}, {Price},
  {Stubbs}, \& {Tonry}}]{Finkbeiner.2016}
{Finkbeiner}, D.~P., {Schlafly}, E.~F., {Schlegel}, D.~J., {et~al.} 2016, \apj,
  822, 66, \dodoi{10.3847/0004-637X/822/2/66}

\bibitem[{{Gerdes} {et~al.}(2016){Gerdes}, {Jennings}, {Bernstein}, {Sako},
  {Adams}, {Goldstein}, {Kessler}, {Hamilton}, {Abbott}, {Abdalla}, {Allam},
  {Benoit-L{\'e}vy}, {Bertin}, {Brooks}, {Buckley-Geer}, {Burke}, {Capozzi},
  {Carnero Rosell}, {Carrasco Kind}, {Carretero}, {Cunha}, {D'Andrea}, {da
  Costa}, {DePoy}, {Desai}, {Dietrich}, {Doel}, {Eifler}, {Fausti Neto},
  {Flaugher}, {Frieman}, {Gaztanaga}, {Gruen}, {Gruendl}, {Gutierrez},
  {Honscheid}, {James}, {Kuehn}, {Kuropatkin}, {Lahav}, {Li}, {Maia}, {March},
  {Martini}, {Miller}, {Miquel}, {Nichol}, {Nord}, {Ogando}, {Plazas}, {Romer},
  {Roodman}, {Sanchez}, {Santiago}, {Schubnell}, {Sevilla-Noarbe}, {Smith},
  {Soares-Santos}, {Sobreira}, {Suchyta}, {Swanson}, {Tarl{\'e}}, {Thaler},
  {Walker}, {Wester}, {Zhang}, \& {DES Collaboration}}]{Gerdes.2016}
{Gerdes}, D.~W., {Jennings}, R.~J., {Bernstein}, G.~M., {et~al.} 2016, \aj,
  151, 39, \dodoi{10.3847/0004-6256/151/2/39}

\bibitem[{{Gladman} {et~al.}(2002){Gladman}, {Holman}, {Grav}, {Kavelaars},
  {Nicholson}, {Aksnes}, \& {Petit}}]{Gladman.2002}
{Gladman}, B., {Holman}, M., {Grav}, T., {et~al.} 2002, \icarus, 157, 269,
  \dodoi{10.1006/icar.2002.6860}

\bibitem[{{Gladman} {et~al.}(2001){Gladman}, {Kavelaars}, {Petit},
  {Morbidelli}, {Holman}, \& {Loredo}}]{Gladman.2001}
{Gladman}, B., {Kavelaars}, J.~J., {Petit}, J.-M., {et~al.} 2001, \aj, 122,
  1051, \dodoi{10.1086/322080}

\bibitem[{{Gladman} \& {Volk}(2021)}]{Gladman.2021}
{Gladman}, B., \& {Volk}, K. 2021, \araa, 59, 203,
  \dodoi{10.1146/annurev-astro-120920-010005}

\bibitem[{{G{\'o}rski} {et~al.}(2005){G{\'o}rski}, {Hivon}, {Banday},
  {Wandelt}, {Hansen}, {Reinecke}, \& {Bartelmann}}]{Gorski.2005}
{G{\'o}rski}, K.~M., {Hivon}, E., {Banday}, A.~J., {et~al.} 2005, \apj, 622,
  759, \dodoi{10.1086/427976}

\bibitem[{{Heinze} {et~al.}(2023){Heinze}, {Eggl}, \& {Juric}}]{Heinze.2023}
{Heinze}, A., {Eggl}, S., \& {Juric}, M. 2023, in AAS/Division for Planetary
  Sciences Meeting Abstracts, Vol.~55, AAS/Division for Planetary Sciences
  Meeting Abstracts \#55, 405.03

\bibitem[{{Heinze} {et~al.}(2022){Heinze}, {Eggl}, {Juric}, {Moeyens}, {Jones},
  {Sullivan}, \& {Bellm}}]{Heinze.2022}
{Heinze}, A., {Eggl}, S., {Juric}, M., {et~al.} 2022, in AAS/Division for
  Planetary Sciences Meeting Abstracts, Vol.~54, AAS/Division for Planetary
  Sciences Meeting Abstracts, 504.04

\bibitem[{{Hodapp} {et~al.}(2004){Hodapp}, {Kaiser}, {Aussel}, {Burgett},
  {Chambers}, {Chun}, {Dombeck}, {Douglas}, {Hafner}, {Heasley}, {Hoblitt},
  {Hude}, {Isani}, {Jedicke}, {Jewitt}, {Laux}, {Luppino}, {Lupton}, {Maberry},
  {Magnier}, {Mannery}, {Monet}, {Morgan}, {Onaka}, {Price}, {Ryan},
  {Siegmund}, {Szapudi}, {Tonry}, {Wainscoat}, \& {Waterson}}]{Hodapp.2004}
{Hodapp}, K.~W., {Kaiser}, N., {Aussel}, H., {et~al.} 2004, Astronomische
  Nachrichten, 325, 636, \dodoi{10.1002/asna.200410300}

\bibitem[{{Holman} {et~al.}(2023){Holman}, {Akmal}, {Farnocchia}, {Rein},
  {Payne}, {Weryk}, {Tamayo}, \& {Hernandez}}]{Holman.2023}
{Holman}, M.~J., {Akmal}, A., {Farnocchia}, D., {et~al.} 2023, \psj, 4, 69,
  \dodoi{10.3847/PSJ/acc9a9}

\bibitem[{{Holman} {et~al.}(2018{\natexlab{a}}){Holman}, {Payne}, {Blankley},
  {Janssen}, \& {Kuindersma}}]{Holman.2018a}
{Holman}, M.~J., {Payne}, M.~J., {Blankley}, P., {Janssen}, R., \&
  {Kuindersma}, S. 2018{\natexlab{a}}, \aj, 156, 135,
  \dodoi{10.3847/1538-3881/aad69a}

\bibitem[{{Holman} {et~al.}(2019){Holman}, {Payne}, \& {P{\'a}l}}]{Holman.2019}
{Holman}, M.~J., {Payne}, M.~J., \& {P{\'a}l}, A. 2019, Research Notes of the
  American Astronomical Society, 3, 160, \dodoi{10.3847/2515-5172/ab4ea6}

\bibitem[{{Holman} {et~al.}(2004){Holman}, {Kavelaars}, {Grav}, {Gladman},
  {Fraser}, {Milisavljevic}, {Nicholson}, {Burns}, {Carruba}, {Petit},
  {Rousselot}, {Mousis}, {Marsden}, \& {Jacobson}}]{Holman.2004}
{Holman}, M.~J., {Kavelaars}, J.~J., {Grav}, T., {et~al.} 2004, \nat, 430, 865,
  \dodoi{10.1038/nature02832}

\bibitem[{{Holman} {et~al.}(2018{\natexlab{b}}){Holman}, {Payne}, {Fraser},
  {Lacerda}, {Bannister}, {Lackner}, {Chen}, {Lin}, {Smith}, {Kokotanekova},
  {Young}, {Chambers}, {Chastel}, {Denneau}, {Fitzsimmons}, {Flewelling},
  {Grav}, {Huber}, {Induni}, {Kudritzki}, {Krolewski}, {Jedicke}, {Kaiser},
  {Lilly}, {Magnier}, {Mark}, {Meech}, {Micheli}, {Murray}, {Parker},
  {Protopapas}, {Ragozzine}, {Veres}, {Wainscoat}, {Waters}, \&
  {Weryk}}]{Holman.2018b}
{Holman}, M.~J., {Payne}, M.~J., {Fraser}, W., {et~al.} 2018{\natexlab{b}},
  \apjl, 855, L6, \dodoi{10.3847/2041-8213/aaadb3}

\bibitem[{{Huang} {et~al.}(2022){Huang}, {Gladman}, {Beaudoin}, \&
  {Zhang}}]{Huang.2022}
{Huang}, Y., {Gladman}, B., {Beaudoin}, M., \& {Zhang}, K. 2022, \apjl, 938,
  L23, \dodoi{10.3847/2041-8213/ac9480}

\bibitem[{{Jester} {et~al.}(2005){Jester}, {Schneider}, {Richards}, {Green},
  {Schmidt}, {Hall}, {Strauss}, {Vanden Berk}, {Stoughton}, {Gunn},
  {Brinkmann}, {Kent}, {Smith}, {Tucker}, \& {Yanny}}]{Jester.2005}
{Jester}, S., {Schneider}, D.~P., {Richards}, G.~T., {et~al.} 2005, \aj, 130,
  873, \dodoi{10.1086/432466}

\bibitem[{{Jones} {et~al.}(2016){Jones}, {Juri{\'c}}, \&
  {Ivezi{\'c}}}]{Jones.2016}
{Jones}, R.~L., {Juri{\'c}}, M., \& {Ivezi{\'c}}, {\v{Z}}. 2016, in Asteroids:
  New Observations, New Models, ed. S.~R. {Chesley}, A.~{Morbidelli},
  R.~{Jedicke}, \& D.~{Farnocchia}, Vol. 318, 282--292,
  \dodoi{10.1017/S1743921315008510}

\bibitem[{{Kaiser} {et~al.}(2010){Kaiser}, {Burgett}, {Chambers}, {Denneau},
  {Heasley}, {Jedicke}, {Magnier}, {Morgan}, {Onaka}, \& {Tonry}}]{Kaiser.2010}
{Kaiser}, N., {Burgett}, W., {Chambers}, K., {et~al.} 2010, in \procspie, Vol.
  7733, Ground-based and Airborne Telescopes III, 77330E,
  \dodoi{10.1117/12.859188}

\bibitem[{{Kavelaars} {et~al.}(2020){Kavelaars}, {Lawler}, {Bannister}, \&
  {Shankman}}]{Kavelaars.2020}
{Kavelaars}, J.~J., {Lawler}, S.~M., {Bannister}, M.~T., \& {Shankman}, C.
  2020, in The Trans-Neptunian Solar System, ed. D.~{Prialnik}, M.~A.
  {Barucci}, \& L.~{Young}, 61--77, \dodoi{10.1016/B978-0-12-816490-7.00003-5}

\bibitem[{{Khain} {et~al.}(2020){Khain}, {Becker}, {Lin}, {Gerdes}, {Adams},
  {Bernardinelli}, {Bernstein}, {Franson}, {Markwardt}, {Hamilton}, {Napier},
  {Sako}, {Abbott}, {Avila}, {Bertin}, {Brooks}, {Buckley-Geer}, {Burke},
  {Carnero Rosell}, {Carrasco Kind}, {Carretero}, {Costa}, {Vicente}, {Desai},
  {Diehl}, {Doel}, {Flaugher}, {Frieman}, {Garc{\'\i}a-Bellido}, {Gaztanaga},
  {Gruen}, {Gruendl}, {Gschwend}, {Gutierrez}, {Hollowood}, {Honscheid},
  {James}, {Kuropatkin}, {Maia}, {Marshall}, {Menanteau}, {Miller}, {Miquel},
  {Plazas}, {Sanchez}, {Scarpine}, {Schubnell}, {Sevilla-Noarbe}, {Smith},
  {Sobreira}, {Suchyta}, {Swanson}, {Tarle}, {Walker}, {Wester}, \& {Dark
  Energy Survey Collaboration}}]{Khain.2020}
{Khain}, T., {Becker}, J.~C., {Lin}, H.~W., {et~al.} 2020, \aj, 159, 133,
  \dodoi{10.3847/1538-3881/ab7002}

\bibitem[{{Kubica} {et~al.}(2007){Kubica}, {Denneau}, {Grav}, {Heasley},
  {Jedicke}, {Masiero}, {Milani}, {Moore}, {Tholen}, \&
  {Wainscoat}}]{Kubica.2007}
{Kubica}, J., {Denneau}, L., {Grav}, T., {et~al.} 2007, \icarus, 189, 151,
  \dodoi{10.1016/j.icarus.2007.01.008}

\bibitem[{{Kurlander} {et~al.}(2025){Kurlander}, {Holman}, {Bernardinelli},
  {Juri{\'c}}, {Heinze}, \& {Payne}}]{Kurlander.2025}
{Kurlander}, J.~A., {Holman}, M.~J., {Bernardinelli}, P.~H., {et~al.} 2025,
  \aj, 169, 73, \dodoi{10.3847/1538-3881/ad9a58}

\bibitem[{{Lawler} {et~al.}(2018){Lawler}, {Kavelaars}, {Alexandersen},
  {Bannister}, {Gladman}, {Petit}, \& {Shankman}}]{Lawler.2018}
{Lawler}, S.~M., {Kavelaars}, J.~J., {Alexandersen}, M., {et~al.} 2018,
  Frontiers in Astronomy and Space Sciences, 5, 14,
  \dodoi{10.3389/fspas.2018.00014}

\bibitem[{{Lin} {et~al.}(2016){Lin}, {Chen}, {Holman}, {Ip}, {Payne},
  {Lacerda}, {Fraser}, {Gerdes}, {Bieryla}, {Sie}, {Chen}, {Burgett},
  {Denneau}, {Jedicke}, {Kaiser}, {Magnier}, {Tonry}, {Wainscoat}, \&
  {Waters}}]{Lin.2016}
{Lin}, H.~W., {Chen}, Y.-T., {Holman}, M.~J., {et~al.} 2016, \aj, 152, 147,
  \dodoi{10.3847/0004-6256/152/5/147}

\bibitem[{{Lykawka} \& {Ito}(2023)}]{Lykawka.2023}
{Lykawka}, P.~S., \& {Ito}, T. 2023, \aj, 166, 118,
  \dodoi{10.3847/1538-3881/aceaf0}

\bibitem[{{Lykawka} \& {Mukai}(2007)}]{Lykawka.2007}
{Lykawka}, P.~S., \& {Mukai}, T. 2007, \icarus, 189, 213,
  \dodoi{10.1016/j.icarus.2007.01.001}

\bibitem[{{Lykawka} \& {Mukai}(2008)}]{Lykawka.2008}
---. 2008, \aj, 135, 1161, \dodoi{10.1088/0004-6256/135/4/1161}

\bibitem[{{Magnier} {et~al.}(2013){Magnier}, {Schlafly}, {Finkbeiner}, {Juric},
  {Tonry}, {Burgett}, {Chambers}, {Flewelling}, {Kaiser}, {Kudritzki},
  {Morgan}, {Price}, {Sweeney}, \& {Stubbs}}]{Magnier.2013}
{Magnier}, E.~A., {Schlafly}, E., {Finkbeiner}, D., {et~al.} 2013, \apjs, 205,
  20, \dodoi{10.1088/0067-0049/205/2/20}

\bibitem[{{Magnier} {et~al.}(2020{\natexlab{a}}){Magnier}, {Schlafly},
  {Finkbeiner}, {Tonry}, {Goldman}, {R{\"o}ser}, {Schilbach}, {Casertano},
  {Chambers}, {Flewelling}, {Huber}, {Price}, {Sweeney}, {Waters}, {Denneau},
  {Draper}, {Hodapp}, {Jedicke}, {Kaiser}, {Kudritzki}, {Metcalfe}, {Stubbs},
  \& {Wainscoat}}]{Magnier.2020a}
{Magnier}, E.~A., {Schlafly}, E.~F., {Finkbeiner}, D.~P., {et~al.}
  2020{\natexlab{a}}, \apjs, 251, 6, \dodoi{10.3847/1538-4365/abb82a}

\bibitem[{{Magnier} {et~al.}(2020{\natexlab{b}}){Magnier}, {Sweeney},
  {Chambers}, {Flewelling}, {Huber}, {Price}, {Waters}, {Denneau}, {Draper},
  {Farrow}, {Jedicke}, {Hodapp}, {Kaiser}, {Kudritzki}, {Metcalfe}, {Stubbs},
  \& {Wainscoat}}]{Magnier.2020b}
{Magnier}, E.~A., {Sweeney}, W.~E., {Chambers}, K.~C., {et~al.}
  2020{\natexlab{b}}, \apjs, 251, 5, \dodoi{10.3847/1538-4365/abb82c}

\bibitem[{{Magnier} {et~al.}(2020{\natexlab{c}}){Magnier}, {Chambers},
  {Flewelling}, {Hoblitt}, {Huber}, {Price}, {Sweeney}, {Waters}, {Denneau},
  {Draper}, {Hodapp}, {Jedicke}, {Kaiser}, {Kudritzki}, {Metcalfe}, {Stubbs},
  \& {Wainscoat}}]{Magnier.2020c}
{Magnier}, E.~A., {Chambers}, K.~C., {Flewelling}, H.~A., {et~al.}
  2020{\natexlab{c}}, \apjs, 251, 3, \dodoi{10.3847/1538-4365/abb829}

\bibitem[{{Maneewongvatana} \& {Mount}(1999)}]{Maneewongvatana.1999}
{Maneewongvatana}, S., \& {Mount}, D.~M. 1999, arXiv e-prints, cs/9901013.
\newblock \doarXiv{cs/9901013}

\bibitem[{{Napier} {et~al.}(2024){Napier}, {Lin}, {Gerdes}, {Adams}, {Simpson},
  {Porter}, {Weber}, {Markwardt}, {Gowman}, {Smotherman}, {Bernardinelli},
  {Juri{\'c}}, {Connolly}, {Kalmbach}, {Portillo}, {Trilling}, {Strauss},
  {Oldroyd}, {Trujillo}, {Chandler}, {Holman}, {Schlichting}, \&
  {McNeill}}]{Napier.2024a}
{Napier}, K.~J., {Lin}, H.~W., {Gerdes}, D.~W., {et~al.} 2024, PSJ, 5, 50,
  \dodoi{10.3847/PSJ/ad1528}

\bibitem[{{Onaka} {et~al.}(2008){Onaka}, {Tonry}, {Isani}, {Lee}, {Uyeshiro},
  {Rae}, {Robertson}, \& {Ching}}]{Onaka.2008}
{Onaka}, P., {Tonry}, J.~L., {Isani}, S., {et~al.} 2008, in \procspie, Vol.
  7014, Ground-based and Airborne Instrumentation for Astronomy II, 70140D,
  \dodoi{10.1117/12.788093}

\bibitem[{{Park} {et~al.}(2021){Park}, {Folkner}, {Williams}, \&
  {Boggs}}]{Park.2021}
{Park}, R.~S., {Folkner}, W.~M., {Williams}, J.~G., \& {Boggs}, D.~H. 2021,
  \aj, 161, 105, \dodoi{10.3847/1538-3881/abd414}

\bibitem[{{Petit} {et~al.}(2006){Petit}, {Holman}, {Gladman}, {Kavelaars},
  {Scholl}, \& {Loredo}}]{Petit.2006}
{Petit}, J.-M., {Holman}, M.~J., {Gladman}, B.~J., {et~al.} 2006, \mnras, 365,
  429, \dodoi{10.1111/j.1365-2966.2005.09661.x}

\bibitem[{{Petit} {et~al.}(2011){Petit}, {Kavelaars}, {Gladman}, {Jones},
  {Parker}, {Van Laerhoven}, {Nicholson}, {Mars}, {Rousselot}, {Mousis},
  {Marsden}, {Bieryla}, {Taylor}, {Ashby}, {Benavidez}, {Campo Bagatin}, \&
  {Bernabeu}}]{Petit.2011}
{Petit}, J.-M., {Kavelaars}, J.~J., {Gladman}, B.~J., {et~al.} 2011, \aj, 142,
  131, \dodoi{10.1088/0004-6256/142/4/131}

\bibitem[{{Rabinowitz} {et~al.}(2012){Rabinowitz}, {Schwamb}, {Hadjiyska}, \&
  {Tourtellotte}}]{Rabinowitz.2012}
{Rabinowitz}, D., {Schwamb}, M.~E., {Hadjiyska}, E., \& {Tourtellotte}, S.
  2012, \aj, 144, 140, \dodoi{10.1088/0004-6256/144/5/140}

\bibitem[{{Schlafly} {et~al.}(2012){Schlafly}, {Finkbeiner}, {Juri{\'c}},
  {Magnier}, {Burgett}, {Chambers}, {Grav}, {Hodapp}, {Kaiser}, {Kudritzki},
  {Martin}, {Morgan}, {Price}, {Rix}, {Stubbs}, {Tonry}, \&
  {Wainscoat}}]{Schlafly.2012}
{Schlafly}, E.~F., {Finkbeiner}, D.~P., {Juri{\'c}}, M., {et~al.} 2012, \apj,
  756, 158, \dodoi{10.1088/0004-637X/756/2/158}

\bibitem[{{Schwamb} {et~al.}(2010){Schwamb}, {Brown}, {Rabinowitz}, \&
  {Ragozzine}}]{Schwamb.2010}
{Schwamb}, M.~E., {Brown}, M.~E., {Rabinowitz}, D.~L., \& {Ragozzine}, D. 2010,
  \apj, 720, 1691, \dodoi{10.1088/0004-637X/720/2/1691}

\bibitem[{{Sheppard} \& {Trujillo}(2016)}]{Sheppard.2016}
{Sheppard}, S.~S., \& {Trujillo}, C. 2016, \aj, 152, 221,
  \dodoi{10.3847/1538-3881/152/6/221}

\bibitem[{{Silsbee} \& {Tremaine}(2018)}]{Silsbee.2018}
{Silsbee}, K., \& {Tremaine}, S. 2018, \aj, 155, 75,
  \dodoi{10.3847/1538-3881/aaa19b}

\bibitem[{{Stern}(1991)}]{Stern.1991}
{Stern}, S.~A. 1991, \icarus, 90, 271, \dodoi{10.1016/0019-1035(91)90106-4}

\bibitem[{{Tonry} \& {Onaka}(2009)}]{PS1_GPCa}
{Tonry}, J., \& {Onaka}, P. 2009, in Advanced Maui Optical and Space
  Surveillance Technologies Conference, E40

\bibitem[{{Tonry} {et~al.}(2012{\natexlab{a}}){Tonry}, {Stubbs}, {Lykke},
  {Doherty}, {Shivvers}, {Burgett}, {Chambers}, {Hodapp}, {Kaiser},
  {Kudritzki}, {Magnier}, {Morgan}, {Price}, \& {Wainscoat}}]{Tonry.2012}
{Tonry}, J.~L., {Stubbs}, C.~W., {Lykke}, K.~R., {et~al.} 2012{\natexlab{a}},
  \apj, 750, 99, \dodoi{10.1088/0004-637X/750/2/99}

\bibitem[{{Tonry} {et~al.}(2012{\natexlab{b}}){Tonry}, {Stubbs}, {Lykke},
  {Doherty}, {Shivvers}, {Burgett}, {Chambers}, {Hodapp}, {Kaiser},
  {Kudritzki}, {Magnier}, {Morgan}, {Price}, \& {Wainscoat}}]{Tonry.2012b}
---. 2012{\natexlab{b}}, \apj, 750, 99, \dodoi{10.1088/0004-637X/750/2/99}

\bibitem[{{Tonry} {et~al.}(2012{\natexlab{c}}){Tonry}, {Stubbs}, {Kilic},
  {Flewelling}, {Deacon}, {Chornock}, {Berger}, {Burgett}, {Chambers},
  {Kaiser}, {Kudritzki}, {Hodapp}, {Magnier}, {Morgan}, {Price}, \&
  {Wainscoat}}]{Tonry.2012a}
{Tonry}, J.~L., {Stubbs}, C.~W., {Kilic}, M., {et~al.} 2012{\natexlab{c}},
  \apj, 745, 42, \dodoi{10.1088/0004-637X/745/1/42}

\bibitem[{{Tonry} {et~al.}(2018){Tonry}, {Denneau}, {Flewelling}, {Heinze},
  {Onken}, {Smartt}, {Stalder}, {Weiland}, \& {Wolf}}]{Tonry.2018}
{Tonry}, J.~L., {Denneau}, L., {Flewelling}, H., {et~al.} 2018, \apj, 867, 105,
  \dodoi{10.3847/1538-4357/aae386}

\bibitem[{{Trilling} {et~al.}(2018){Trilling}, {Bellm}, \&
  {Malhotra}}]{Trilling.2018}
{Trilling}, D.~E., {Bellm}, E.~C., \& {Malhotra}, R. 2018, \aj, 155, 243,
  \dodoi{10.3847/1538-3881/aabfc0}

\bibitem[{{Trilling} {et~al.}(2024){Trilling}, {Gerdes}, {Juri{\'c}},
  {Trujillo}, {Bernardinelli}, {Napier}, {Smotherman}, {Strauss}, {Fuentes},
  {Holman}, {Lin}, {Markwardt}, {McNeill}, {Mommert}, {Oldroyd}, {Payne},
  {Ragozzine}, {Rivkin}, {Schlichting}, {Sheppard}, {Adams}, \&
  {Chandler}}]{Trilling.2024}
{Trilling}, D.~E., {Gerdes}, D.~W., {Juri{\'c}}, M., {et~al.} 2024, \aj, 167,
  132, \dodoi{10.3847/1538-3881/ad1529}

\bibitem[{{Trujillo} {et~al.}(2001){Trujillo}, {Jewitt}, \&
  {Luu}}]{Trujillo.2001}
{Trujillo}, C.~A., {Jewitt}, D.~C., \& {Luu}, J.~X. 2001, \aj, 122, 457,
  \dodoi{10.1086/321117}

\bibitem[{{Trujillo} \& {Sheppard}(2014)}]{Trujillo.2014}
{Trujillo}, C.~A., \& {Sheppard}, S.~S. 2014, \nat, 507, 471,
  \dodoi{10.1038/nature13156}

\bibitem[{{Volk} \& {Malhotra}(2017)}]{Volk.2017}
{Volk}, K., \& {Malhotra}, R. 2017, \aj, 154, 62,
  \dodoi{10.3847/1538-3881/aa79ff}

\bibitem[{{Weryk} {et~al.}(2016){Weryk}, {Lilly}, {Chastel}, {Denneau},
  {Jedicke}, {Magnier}, {Wainscoat}, {Chambers}, {Flewelling}, {Huber},
  {Waters}, \& {PS1 Builders}}]{Weryk.2016}
{Weryk}, R.~J., {Lilly}, E., {Chastel}, S., {et~al.} 2016, arXiv e-prints,
  arXiv:1607.04895.
\newblock \doarXiv{1607.04895}

\bibitem[{Zonca {et~al.}(2019)Zonca, Singer, Lenz, Reinecke, Rosset, Hivon, \&
  Gorski}]{Zonca.2019}
Zonca, A., Singer, L., Lenz, D., {et~al.} 2019, Journal of Open Source
  Software, 4, 1298, \dodoi{10.21105/joss.01298}

\end{thebibliography}

\end{document}